\documentclass[12pt]{article}

\usepackage{graphicx}
\usepackage{amsfonts}
\usepackage{amssymb,amsmath,amscd}
\usepackage[top=2.75cm, bottom=2.75cm, left=2.75cm, right=2.75cm]{geometry} 

\newtheorem{proposition}{Proposition}

\newtheorem{definition}[proposition]{Definition}

\newtheorem{theorem}[proposition]{Theorem}

\def\+{{+\!\!\!+}}

\def\d{\partial}

\def\pmb#1{\setbox0=\hbox{#1}%
\kern.0em\copy0\kern-\wd0 
\kern-.04em\copy0\kern-\wd0 
\kern.08em\copy0\kern-\wd0 
\kern-.04em\raise.0433em\box0 }         
 
\newcommand{\nc}{\newcommand} 
\nc{\beq}{\begin{equation}} 
\nc{\eeq}[1]{\label{#1}\end{equation}} 
\nc{\ber}{\begin{eqnarray}} 
\nc{\eer}[1]{\label{#1}\end{eqnarray}} 
\nc{\pek}[1]{\cite{#1}} 
\nc{\enr}[1]{(\ref{#1})} 
\nc{\kal}[1]{{\cal{#1}}} 
\nc{\dott}{\;\cdot\;} 
\nc{\coker}{\mathrm{coker}}
\nc{\ie}{{\it i.e.}}
\nc{\eg}{{\it e.g.}}
\newcommand{\Section}[1]{\section{#1} \setcounter{equation}{0}}

\def\0 {\nonumber}

\begin{document} 
\setcounter{page}{0}
\newcommand{\inv}[1]{{#1}^{-1}} 
\renewcommand{\theequation}{\thesection.\arabic{equation}} 
\newcommand{\be}{\begin{equation}} 
\newcommand{\ee}{\end{equation}} 
\newcommand{\bea}{\begin{eqnarray}} 
\newcommand{\eea}{\end{eqnarray}} 
\newcommand{\re}[1]{(\ref{#1})} 
\newcommand{\qv}{\quad ,} 
\newcommand{\qp}{\quad .} 

\thispagestyle{empty}
\begin{flushright} \small
UUITP-06/07  \\
\end{flushright}
\smallskip
\begin{center} \LARGE
{\bf Poisson sigma model on the sphere}
 \\[12mm] \normalsize
{\large\bf Francesco Bonechi$^{a}$ and Maxim Zabzine$^{b}$} \\[8mm]
 {\small\it
$^a$I.N.F.N. and Dipartimento di Fisica,\\
  Via G. Sansone 1, 50019 Sesto Fiorentino - Firenze, Italy \\
~\\
$^b$Department of Theoretical Physics 
Uppsala University, \\ Box 803, SE-751 08 Uppsala, Sweden
\\~\\
}
\end{center}
\vspace{10mm}
\centerline{\bfseries Abstract} \bigskip
 We evaluate the path integral of the Poisson sigma model on sphere and
  study  the correlators of quantum observables. 
    We argue that for the path integral to be well-defined the corresponding 
   Poisson structure should be unimodular.    
   The construction of  the finite dimensional BV theory is presented 
    and we argue that it is responsible for the leading semiclassical contribution. 
    For a (twisted) generalized
     K\"ahler manifold we  discuss the gauge fixed action for the Poisson sigma model.
      Using the localization we prove that for the holomorphic Poisson structure the semiclassical 
       result for the correlators is indeed  the full quantum result.

\noindent  

\eject
\normalsize



\section{Introduction}
\label{start}

The Poisson sigma model (PSM), introduced in \cite{Ikeda:1993fh, Schaller:1994es}, is a topological
 two-dimensional field theory  with target a Poisson manifold $M$, whose
  Poisson tensor we will denote by $\alpha$ throughout.  Recently PSM has attracted a lot of attention 
   due to its role in the deformation quantization \cite{Cattaneo:1999fm}.  In particular 
    the star product is given by a semiclassical expansion of the path integral of the PSM over the disk. 
     In the present paper we study the PSM defined over the sphere.  
  
  Let us start with a brief reminder of PSM. 
Take  $\Sigma$ to be a two-dimensional oriented compact manifold
 without boundary. The starting point is the classical action functional $S$
 defined on the space of vector bundle morphisms $\hat{X}\colon  T\Sigma \rightarrow T^*M$
 from the tangent bundle $T\Sigma$ to the cotangent
 bundle $T^*M$ of the Poisson manifold $M$.
 Such a map $\hat{X}$ is given by its base map $X\colon  \Sigma \rightarrow M$  and the linear map $\eta$ between
 fibers, which may also be regarded as
 a section in $\Gamma(\Sigma, Hom(T\Sigma, X^*(T^*M)))$.
  The pairing $\langle\,\,,\,\,\rangle$ between the cotangent and tangent space at each point of $M$
  induces a pairing between the differential forms on $\Sigma$ with values in the pull-backs
 $X^*(T^*M)$ and $X^*(TM)$ respectively. It is defined
 as pairing of the values and the exterior product of differential forms.
 Then the action functional $S$ of the theory is
\begin{equation} S(X,\eta) = \int\limits_\Sigma  \langle \eta,  dX\rangle +
\frac{1}{2} \langle \eta, (\alpha \circ X) \eta \rangle~ .
\label{definPS}
\end{equation}
 Here $\eta$ and $dX$ are viewed as one-forms on $\Sigma$ with the values in the pull-back of
 the cotangent and tangent bundles of $M$ correspondingly.
 Thus, in local coordinates, we can rewrite the action (\ref{definPS}) as follows:
 \begin{equation}
 S(X,\eta) = \int\limits_D  \eta_\mu \wedge dX^\mu + \frac{1}{2} \alpha^{\mu\nu}(X) \eta_\mu
 \wedge \eta_\nu~ .
\label{definPSloccoor}\end{equation}
 The variation of the action gives rise to the following equations of
 motion
\begin{equation}
 d\eta_\rho + \frac{1}{2} (\d_\rho \alpha^{\mu\nu}) \eta_\mu \wedge \eta_\nu =0~,~~~~~~~
  dX^\mu + \alpha^{\mu\nu}\eta_\nu = 0~ .
\label{eqmotion}\end{equation}
 In covariant language these equations are equivalent to the statement that
 the bundle morphism $\hat{X}$ is  a Lie algebroid morphism  from
 $T\Sigma$ (with standard Lie algebroid structure) to $T^*M$ (with Lie algebroid structure canonically
 induced by the Poisson structure).
The action (\ref{definPSloccoor}) is invariant under the infinitesimal gauge transformations
\begin{equation}
 \delta_\beta X^\mu = \alpha^{\mu\nu} \beta_\nu,\,\,\,\,\,\,\,\,\,\,\,\,\,
 \delta_\beta \eta_\mu = - d\beta_{\mu} - (\d_\mu \alpha^{\nu\rho}) \eta_\nu \beta_\rho~ ,
\label{gaugetransf}\end{equation}
 which form a closed algebra only on-shell (i.e., modulo the equations of motion (\ref{eqmotion})).
 
 In order to quantize the PSM we have to resolve to the Batalin-Vilkovisky 
  (BV) formalism \cite{Batalin:1977pb} which we will review later.   In what follows we will be concentrated mainly on the case when the world-sheet $\Sigma$  is
   two-sphere $S^2$.  Our goal is to calculate a leading term for PSM correlators on $S^2$.
     We will argue that the notion of unimodularity appear naturally in the construction of the correlators.
 Indeed  our construction is very similar to the one presented in \cite{Pestun:2006rj} and 
  is a generalization of the correlators for A- and B-models (see \cite{Hori:2003ic} for review). 
  It is not surprising since the notion of generalized Calabi-Yau manifold given in \cite{hitchinCY} 
   is a complex version of 
   the notion of unimodularity of a Lie algebroid.  In particular the unimodularity of  Poisson manifold is 
    a real analog of  generalized Calabi-Yau condition.  Previously in the different context the path integral for PSM and related models
     was also discussed in \cite{Kummer:1996hy, Hirshfeld:2001cm, Bergamin:2004pn}.
   
   In the second part of the paper we consider a particular gauge fixing which involves a choice of an
    (almost) complex structure. The whole setup is realized on (twisted) generalized K\"ahler manifolds. 
     For these gauge fixed models there exists a residual BRST symmetry which allows to 
      use the localization. Thus we are able to produce examples where the leading term is 
       a full answer for the quantum theory.  
 
 The paper is organized as follows. In Section \ref{reviewBV} we review basic concepts of BV formalism.
   Section \ref{BV} is devoted to overview of BV treatment of PSM. In particular we discuss
    the classical observables.  In Section \ref{multivectors} we consider the truncation of the full 
     BV theory to a finite dimensional BV theory which is responsible for the leading semiclassical contribution in 
      the correlators.  We discuss this finite dimensional BV theory in details.  In this context the unimodularity of Poisson manifold  arises naturally from the quantum master equation. 
    In Section \ref{gauge} the specific gauge fixing is discussed. Indeed the geometrical set-up 
   we are using   is the same as for the $N=2$ supersymmetric PSM \cite{Calvo:2005ww}. We work out 
    the details of gauge fixing and discuss the residual BRST transformations of the gauge fixed 
     action and present the calculations of the correlators for the gauge fixed model. 
       Finally Section \ref{end} summarizes the results and discusses open issues. 
 
  In addition we have  Appendices A and B where the relevant mathematical 
   material is collected. The material presented there is not entirely  original and furthermore  we
    could not find appropriate  references with all material. 
   Many of the results presented in Appendices are scattered throughout the literature. 
    Moreover we would like to link two different languages used by different communities. 
       In particular   the notion of generalized Calabi-Yau manifold introduced 
    by Hitchin \cite{hitchinCY} is related to the notion of unimodularity for complex 
     Lie algebroid.  

 Throughout the paper we use the language of graded manifolds which are
supermanifolds with a $\mathbb{Z}$-refinement of $\mathbb{Z}_2$-grading, e.g.
    see \cite{roytenberg1} for the review.

\section{Review of BV formalism}
\label{reviewBV}

 In this Section we briefly review the relevant concepts within the general BV framework.
 For further details the reader may consult the following
   reviews \cite{cattaneobv, fiorenza, henneaux}. 
  
\begin{definition}\label{Gerstenhaberalg}
A graded algebra ${\cal A}$ with an odd bracket $\{~,~\}$ is called an odd Poisson algebra (Gerstenhaber
 algebra) if the bracket satisfies 
 $$ \{ f, g\} = -(-1)^{(|f|+1)(|g|+1)}\{g, f\}~,$$
 $$ \{f, \{g, h\}\} =\{\{f,g\},h\} + (-1)^{(|f|+1)(|g|+1)}\{g,\{f,h\}\}~,$$
 $$\{ f, gh\} = \{f, g\}h + (-1)^{(|f|+1)|g|} g\{f,h\}~.$$
\end{definition}
  Quite often such odd Poisson bracket is called either  Gerstenhaber bracket or antibracket.
  
 \begin{definition}\label{BValgebra}
 A Gerstenhaber algebra $({\cal A}, \{~,~\})$ together with an odd $\mathbb{R}$--linear map
 $$\Delta ~:~{\cal A} \longrightarrow {\cal A}\;,$$ 
  which squares to zero $\Delta^2=0$ and generates the bracket $\{~,~\}$ as
   $$\{f, g\}=  (-1)^{|f|} \Delta(fg)+(-1)^{|f|+1}(\Delta f)g - f(\Delta g) \;,$$
  is called a BV-algebra. $\Delta$ is called odd Laplace operator (odd Laplacian). 
 \end{definition}
The canonical example of BV algebra is given by the space of functions on $W\oplus \Pi W^*$,
where $W$ is a superspace, $W^*$ is its dual and $\Pi$ stands for the reversed parity functor.  
$W\oplus \Pi W^*$ is equipped with an odd non-degenerate pairing. Let $y^a$ be the coordinates
on $W$ (the fields) and $y_a^+$ be the corresponding coordinates 
on $\Pi W^*$ (the antifields). We denote the parity of $y^a$ as $(-1)^{|y^a|}$ and that of $y^+_a$ as $(-1)^{|y^+_a|}=(-1)^{|y^a|+1}$. Then the odd Laplacian is defined as follows
     \beq
      \Delta = (-1)^{|y_a|}\frac{\d}{\d y_a^+}\frac{\d}{\d y^a}~.
     \eeq{trivialoddlap}
 It generates the canonical antibracket on $C^\infty ( W\oplus \Pi W^*)$
 \beq
  \{ f, g\} = 
   (-1)^{|y^a|}\frac{\overleftarrow{\d}f}{\d y_a^+} \frac{\overrightarrow{\d}g}{\d y^a} +
   (-1)^{|y^a|}\frac{\overleftarrow{\d}f}{\d y^a} \frac{\overrightarrow{\d}g}{\d y_a^+}
   ~,
 \eeq{generalcxlsoo}
  where we use the notation $\overrightarrow{\d}_v f = \d_v f$ and $\overleftarrow{\d}_v f = (-1)^{|v||f|}
   \d_v f$.  Indeed the bracket (\ref{generalcxlsoo}) is non degenerate and defines the canonical odd symplectic structure on $W\oplus \Pi W^*$.  
   
   A Lagrangian submanifold ${\cal L} \subset W\oplus \Pi W^*$ is an isotropic supermanifold of maximal 
    dimension. The volume form $dy^1...dy^n dy_1^+...dy_n^+$ induces a well defined volume form 
     on ${\cal L}$. Thus the integral 
     \beq
       \int\limits_{\cal L} f,~~~~~~~f\in C^\infty(W\oplus \Pi W^*)
     \eeq{definelagen}
 is defined for any ${\cal L}$.  The following is the main theorem of BV-formalism.
 
 \begin{theorem}\label{BVtheorem}
  If $\Delta f =0$, then $\int\limits_{\cal L} f$ depends only on the homology class of ${\cal L}$. Moreover $\int\limits_{\cal L} \Delta f =0$ for any Lagrangian ${\cal L}$. 
 \end{theorem}
  
 The canonical example $W\oplus \Pi W^*$ can be generalized to the
cotangent bundle $T^*[-1] {\cal M}$ of any graded manifold ${\cal M}$ \cite{Schwarz:1992nx}. As a cotangent bundle, $T^*[-1] {\cal M}$ is naturally equipped with an odd Poisson bracket that makes  
$C^\infty (T^*[-1] {\cal M})$ a Gerstenhaber algebra according to Definiton \ref{Gerstenhaberalg}.
The idea is that locally one can map $T^*[-1] {\cal M}$ to $W\oplus \Pi W^*$, define the bracket on coordinates with (\ref{generalcxlsoo}) and then glue the patches in a consistent manner. 
    
 Now in order to  define the odd Laplacian $\Delta$ we need an integration over $T^*[-1] {\cal M}$. 
 Namely, the choice of a volume form $v$ on ${\cal M}$ produces the corresponding volume form $\mu_v$ on $T^*[-1]{\cal M}$. The divergence operator is defined as a map
     from the vector fields on $T^*[-1]{\cal M}$ to $C^\infty(T^*[-1]{\cal M})$ through the 
    following  integral relation  
   \beq
  \int\limits_{T^*[-1]{\cal M}}   X(f) ~\mu_v =   - \int\limits_{T^*[-1]{\cal M}}  {\rm div}_{\mu_v} X ~ f ~\mu_v~,~~~~~~~~
  \forall f \in C^\infty (T^*[-1] {\cal M})~,
 \eeq{divergencedef}
 with $X$ being a vector field. As one can easily check, for any function $f$ and vector field $X$ the divergence satisfies 
\begin{equation}\label{divergence}
 {\rm div}_{\mu_v}(f X)= f {\rm div}_{\mu_v}(X) + (-1)^{|f||X|} X(f)\;.
\end{equation}

Now the odd Laplacian of $f\in C^\infty(T^*[-1]\cal M)$ is defined through the divergence of the corresponding Hamiltonian vector field as
  \beq
   \Delta_v f = \frac{(-1)^{|f|}}{2}{\rm div_{\mu_v}} X_f~,~~~~~\{f, g\} =X_f(g)~.
  \eeq{definlapppekdk}
 Indeed one can check that thanks to (\ref{divergence}) $\Delta_v$ generates the bracket and $\Delta_v^2=0$.  Thus 
  $C^\infty (T^*[-1] {\cal M})$ is a BV-algebra according to Definition \ref{BValgebra}, see \cite{kosmann3}
   for the explicit calculations. If the volume form is written in terms of an even density $\rho_v$ as 
   $$\mu_v=\rho_v dy^1\cdots dy^n dy^+_1\cdots dy^+_n~,$$ 
   then the Laplacian can be written as
  \begin{equation}\label{global_laplacian}
  \Delta_v= (-1)^{|y_a|}\frac{\d}{\d y_a^+}\frac{\d}{\d y^a} + \frac{1}{2} \{\log\rho_v, -\}~.
  \end{equation} 

 There exists a canonical way (up to a sign) of restricting a volume form $\mu_v$ on 
  $T^*[-1] {\cal M}$ to a volume form on a Lagrangian submanifold ${\cal L}$. We denote 
   such restriction as $\sqrt{\mu_v}$ and consider  the integrals of the form
  \beq
   \int\limits_{\cal L} \sqrt{\mu_v}~ f~,~~~~~~~~ f \in C^\infty (T^*[-1] {\cal M})~.
  \eeq{integhajw28920}  
Thus the Theorem \ref{BVtheorem} will remain to be true for the general case. 
 In particular we are interested in the situation when the integrands in (\ref{integhajw28920}) are of the form
\beq
 \int\limits_{\cal L}  \sqrt{\mu_v} ~\Psi e^{S} \equiv \langle \Psi \rangle~,
\eeq{confjdjdkkkll} 
 where we assume naturally that $\Delta_v (\Psi e^{S})=0$. 
If $\Psi=1$ then we get the following relation
\beq
 \Delta_v \left ( e^S \right ) =0~~ \Longleftrightarrow~~ \Delta_v S + \frac{1}{2} \{S, S\}=0~,
\eeq{rela327388}
 which is known as the {\it quantum master equation}. In the general case we have
\beq
\Delta_v \left ( \Psi e^S \right ) =0~~ \Longleftrightarrow~~   \Delta_{(v,S)} \Psi = \Delta_v \Psi + \{ S , \Psi\}=0~,
\eeq{definitionquantum}
 where we refer to $\Delta_{(v,S)}$ as the quantum Laplacian.  In the derivation 
  of (\ref{definitionquantum}) we have used the quantum master equation (\ref{rela327388}). A function
   $S$ that satisfies the quantum master equation is called a quantum  BV action and $\Psi$ satisfying (\ref{definitionquantum}) is a quantum observable. Indeed the quantum observables are elements of the
cohomology $H(\Delta_{(v,S)})$; by the above construction it is clear that $S$ defines the isomorphism
  \beq
   H^\bullet (\Delta_v) \approx H^\bullet(\Delta_{(v,S)})\;.
  \eeq{isomrocohomls;}
 
 If we change $S$ to $S/\hbar$, we see that in the classical limit ($\hbar \rightarrow 0$) $S$ must satisfy the classical master equation $\{ S, S\}=0$ and the classical observables $\Psi$ are such that $\delta_{BV} \Psi \equiv \{ S, \Psi \}=0$. Due to the classical master equation the vector field $\delta_{BV}$ squares to zero and defines the cohomology $H(\delta_{BV})$ of classical observables.

 If ${\cal M}$ is a finite dimensional manifold then everything is well-defined. However in field theory 
  one deals with ${\cal M}$ being infinite dimensional. 
 In fact, $\cal M$ is usually the space of the physical fields, ghosts and Lagrange multipliers, that is infinite dimensional. We extend this set of fields by adding antifields such that together they form $T^*[-1]{\cal M}$, where an odd Poisson bracket is well-defined on large enough class of functions, as described above.
 However there is no well-defined measure on ${\cal M}$ and thus there is no well-defined odd Laplace operators. In physics literature, the naive Laplacian of the form  (\ref{generalcxlsoo}) is used. 
 Moreover the field theory suffers from the problems with  renormalization which can be resolved within the perturbative setup. 

\section{BV formalism for PSM}
\label{BV}

The quantization of PSM requires the machinery of BV formalism. 
 In this Section we set the notation and give a background information  on the BV treatment of PSM. 
  We mainly review the relevant results from from \cite{Cattaneo:1999fm} and \cite{Cattaneo:2001ys}.
   Furthermore we discuss the classical observables. 

\subsection{BV action}
\label{BVaction}

The PSM action (\ref{definPSloccoor}) 
 has gauge symmetries which  do not close off-shell. Therefore one 
 should resort to BV formalism. We may organize the fields, ghosts and antifields into superfields
  $({\mathsf X}, \boldsymbol{\eta})$ which corresponds to the components of supermap 
   $T[1]\Sigma \rightarrow T^*[1]M$. Introducing the local coordinates on $\Sigma$ and $M$
 the superfields read as
$$ {\mathsf X}^\mu = X^\mu + \theta^\alpha \eta_\alpha^{+\mu} - \frac{1}{2} \theta^\alpha\theta^\beta
\beta^{+\mu}_{\alpha\beta}~,$$
$$ \boldsymbol{\eta}_\mu = \beta_\mu + \theta^\alpha \eta_{\alpha\mu} + \frac{1}{2} \theta^\alpha\theta^\beta X^{+}_{\alpha\beta\mu}~,$$
  with $\theta$ being the odd coordinate on $\Pi T\Sigma$, $\alpha,\beta$ are labels for local coordinates
   on $\Sigma$ and $\mu$ are labels for local coordinates on $M$. 
In the expansion 
  $\beta$ is a ghost with the ghost number $1$, while $\eta^+$, $\beta^+$ and $X^+$ are 
  antifields of ghost number $-1$, $-2$ and $-1$ respectively.  The full BV action reads
  \beq
   S_{BV} = \int d^2\theta d^2 u\, \left ( \boldsymbol{\eta}_\mu D {\mathsf X}^\mu +
    \frac{1}{2} \alpha^{\mu\nu}({\mathsf X}) \boldsymbol{\eta}_\mu \boldsymbol{\eta}_\nu\right )~,
  \eeq{fullBVsuperfiel}
 where $D= \theta^\alpha \d_\alpha$. An elegant way to derive this action is to use the AKSZ formalism 
  \cite{Alexandrov:1995kv} as done in \cite{Cattaneo:2001ys}. 
  On $T^*[-1]{\cal M}$ the odd symplectic structure is 
   \beq
    \omega = \int\limits_{\Sigma} \left ( \delta X \wedge \delta X^+ + \delta \eta\wedge \delta \eta^+ +
     \delta \beta \wedge \delta \beta^+ \right )~,  
   \eeq{oddsymplecticsty}
    where ${\cal M}$ is infinite dimensional manifold corresponding to the fields $(X,\eta,\beta)$.
   The action (\ref{fullBVsuperfiel})
   satisfies both classical and naive quantum master equations \cite{Cattaneo:1999fm}. 
 The corresponding BRST operator $\delta_{BV}$ acts on the superfields as follows
\ber
\label{EQ123}&&\delta_{BV} {\mathsf X}^\mu = D{\mathsf X}^\mu + \alpha^{\mu\nu} ({\mathsf X}) \boldsymbol{\eta}_\nu~ ,\\ 
\label{EQ1234}&& \delta_{BV} \boldsymbol{\eta}_\mu = D\boldsymbol{\eta}_\mu + \frac{1}{2} \d_\mu \alpha^{\nu\rho}({\mathsf X}) \boldsymbol{\eta}_\nu \boldsymbol{\eta}_\rho~ .
\eer{BRSTinsuperfield}
 In component the BV action  (\ref{fullBVsuperfiel}) has the form
$$S_{BV} = \int\limits_{\Sigma} \eta_\mu \wedge dX^\mu + \frac{1}{2} \alpha^{\mu\nu} (X)
 \eta_\mu \wedge \eta_\nu + X^+_\mu \alpha^{\mu\nu}(X) \beta_\nu - \eta^{+\mu} \wedge \left (d\beta_\mu
  + \d_\mu \alpha^{\rho\nu}(X) \eta_\rho \beta_\nu\right ) - $$
  \beq
  -\frac{1}{2} \beta^{+\mu} \partial_\mu \alpha^{\rho\nu}(X)
  \beta_\rho\beta_\nu - \frac{1}{4} \eta^{+\mu} \wedge \eta^{+\nu} \partial_\mu \partial_\nu \alpha^{\rho\sigma} (X)\beta_\rho \beta_\sigma~ . 
\eeq{BVfullaction}
The component version of the BV transformations (\ref{EQ123})-(\ref{EQ1234}) is
\ber
\label{AAAA1}&&\delta_{BV} X^\mu = \alpha^{\mu\nu}(X)\beta_\nu~,\\
\label{AAAA2}&&\delta_{BV} \eta^{+\mu} = - dX^\mu  - \alpha^{\mu\nu} (X)\eta_\nu - \d_\nu \alpha^{\mu\rho}(X) \eta^{+\nu} \beta_\rho~,\\
\label{AAAA3}&&\nonumber  \delta_{BV} \beta^{+\mu} = - d\eta^{+\mu} - \alpha^{\mu\nu}(X) X_\nu^+ + \frac{1}{2} \d_\nu \d_\rho \alpha^{\mu\sigma}(X) \eta^{+\nu} \wedge \eta^{+\rho} \beta_\sigma + \\
\label{AAAA4}&&\,\,\,\,\,\,\,\,\,\,\,\,\,\,\,\,\,\,\,\,\,+\d_\rho \alpha^{\mu\nu} (X)\eta^{+\rho} \wedge \eta_\nu + \d_\rho \alpha^{\mu\nu}(X) \beta^{+\rho} \beta_\nu~,\\
\label{AAAA5}&&\delta_{BV} \beta_\mu = \frac{1}{2} \d_\mu \alpha^{\nu\rho}(X) \beta_{\nu} \beta_\rho~,\\
\label{AAAA6}&& \delta_{BV} \eta_\mu = - d\beta_\mu - \d_\mu \alpha^{\nu\rho}(X) \eta_\nu \beta_\rho - \frac{1}{2}
\d_\mu \d_\nu \alpha^{\rho\sigma}(X) \eta^{+\nu} \beta_\rho \beta_\sigma ~,\\
&&\nonumber  \delta_{BV} X^+_\mu = d\eta_\mu + \d_\mu \alpha^{\nu\rho}(X) X_\nu^+ \beta_\rho -
\d_\mu \d_\nu \alpha^{\rho\sigma}(X) \eta^{+\nu} \wedge \eta_\rho \beta_\sigma +
 \frac{1}{2} \d_\mu \alpha^{\nu\rho}(X) \eta_\nu \wedge \eta_\rho -\\
\label{AAAA7} &&\,\,\,\,\,\,\,\,\,\,\,\,\,\,\,\,\,\,\,\,-\frac{1}{4} \d_\mu \d_\nu \d_\rho \alpha^{\sigma\tau}(X) \eta^{+\nu} \wedge \eta^{+\rho}
 \beta_\sigma \beta_\tau - \frac{1}{2} \d_\mu\d_\nu \alpha^{\rho\sigma}(X) \beta^{+\nu} \beta_\rho
 \beta_{\sigma}~.
\eer{hhdhdjal}

\subsection{Classical observables}
\label{observables}

Next we consider the classical observables for PSM. By an observable we mean a BRST invariant 
 operator which is not BRST exact. 

Let us take antisymmetric multivector field $w \in \Gamma( \wedge^p TM)$ and construct 
 the superfield $w^{{\mu_1}...{\mu_p}}({\mathsf X})\boldsymbol{\eta}_{\mu_1} ...\boldsymbol{\eta}_{\mu_p}$.
 Using (\ref{EQ123})-(\ref{EQ1234}) we calculate 
  the BRST transformation of this superfield
\beq
 \delta_{BV} (w^{{\mu_1}...{\mu_p}}\boldsymbol{\eta}_{\mu_1} ...\boldsymbol{\eta}_{\mu_p}) = D(w^{{\mu_1}...{\mu_p}}\boldsymbol{\eta}_{\mu_1} ...\boldsymbol{\eta}_{\mu_p})  - \frac{1}{2}( [\alpha, w]_s)^{{\mu_0}{\mu_1}...{\mu_p}} \boldsymbol{\eta}_{\mu_0} \boldsymbol{\eta}_{\mu_1} ...\boldsymbol{\eta}_{\mu_p}~ .
\eeq{caldle039399}
 The last term on the right hand side vanishes if $d_{LP} w= [\alpha, w]_s=0$. Moreover we do not
  want the superfield $w^{{\mu_1}...{\mu_p}}\boldsymbol{\eta}_{\mu_1} ...\boldsymbol{\eta}_{\mu_p}$ to be BRST exact. Thus we have to take $w$ to be an element in the Lichnerowicz-Poisson cohomoogy
   $H^\bullet_{LP}(M)$.  Now assuming $[w] \in H^\bullet_{LP}(M)$ 
   we can interpret (\ref{caldle039399}) in components. The superfield has the expansion
$$ w^{{\mu_1}...{\mu_p}}\boldsymbol{\eta}_{\mu_1} ...\boldsymbol{\eta}_{\mu_p}= O_0^p 
 +\theta^\alpha (O_1^{p-1})_\alpha + \frac{1}{2} \theta^\alpha \theta^\beta (O_2^{p-2})_{\alpha\beta}$$
 on which the BRST differential $\delta_{BV}$ acts as
$$ \delta_{BV} (w^{{\mu_1}...{\mu_p}}\boldsymbol{\eta}_{\mu_1} ...\boldsymbol{\eta}_{\mu_p})= \delta_{BV} O_0^p 
 -\theta^\alpha \delta_{BV} (O_1^{p-1})_\alpha + \frac{1}{2} \theta^\alpha \theta^\beta \delta_{BV}(O_2^{p-2})_{\alpha\beta}~.$$
  The operator $D=\theta^\alpha\d_\alpha$ acts on the component fields as the de Rham differential. 
   Thus for $[w] \in H^\bullet_{LP}(M)$ the condition (\ref{caldle039399}) implies
     the descent equations for the components
    \beq
    \delta_{BV} O_0^p=0~,\,\,\,\,\,\,\,\,\,\,\,\,\,\,\,\,\,\,
    \delta_{BV} O_1^{p-1} = - d Q_0^p~,\,\,\,\,\,\,\,\,\,\,\,\,\,\,\,\,\,\,
    \delta_{BV} O_2^{p-2} = d Q_1^{p-1}~. 
    \eeq{descenteqowo}
More explicitly  for a nontrivial element 
 $[w]\in H^p_{LP}(M)$ we can formally define the cocycles
 \ber
 && O_0^p (w) = w^{{\mu_1}...{\mu_p}} \beta_{\mu_1}...\beta_{\mu_p}~,\\
 && O_1^{p-1} (w) = \d_\rho w^{{\mu_1}...{\mu_p}} \eta^{+\rho} \beta_{\mu_1}...\beta_{\mu_p}
  + p w^{{\mu_1}{\mu_2}...{\mu_p}} \eta_{\mu_1}\beta_{\mu_2}...\beta_{\mu_p}~,\\
 \nonumber && O_2^{p-2} (w) = -\frac{1}{2} \d_\rho\d_\sigma w^{{\mu_1}...{\mu_p}} \eta^{+\rho} \wedge \eta^{+\sigma}
  \beta_{\mu_1}...\beta_{\mu_p} - \d_\rho w^{{\mu_1}...{\mu_p}} \beta^{+\rho}\beta_{\mu_1}...\beta_{\mu_p} -\\
  \nonumber&&\,\,\,\,\,\,\,\,\,\,\,\,\,\,\,\,\,\,\,\,\,- p \d_\rho w^{{\mu_1}...{\mu_p}} \eta^{+\rho} \wedge \eta_{\mu_1}\beta_{\mu_2}...\beta_{\mu_p}
  + p w^{{\mu_1}...{\mu_p}} X^+_{\mu_1}\beta_{\mu_2}...\beta_{\mu_p} + \\
  \label{38383iiiskkka}&&\,\,\,\,\,\,\,\,\,\,\,\,\,\,\,\,\,\,\,\,\,+p(p-1) w^{{\mu_1}...{\mu_p}}
  \eta_{\mu_1}\wedge \eta_{\mu_2}\beta_{\mu_3}...\beta_{\mu_p}~,
 \eer{obser2789390}
  where in $O_i^{p-i}(w)$ the upper index stands for the ghost number and the lower index for the 
   degree of the differential form on $\Sigma$. $Q_i^{p-i}(w)$ satisfy (\ref{descenteqowo}) and
      thus $O_0^p(w)$ are BRST-invariant local observables labeled by the elements of the Lichnerowicz-Poisson cohomology $H^\bullet_{LP}(M)$. From $O_i^{p-i}(w)$ with $i>0$ we can construct BRST-invariant 
    non-local observables as integrals
    \beq
  W(w , c_i)=   \int\limits_{c_i} O_i^{p-i} (w)
    \eeq{indetaiiejj}
    where $c_i$ is $i$-cycle on $\Sigma$. These observables depend only on the homology class of $c_i$. The antibracket $\{~,~\}$ of two non-local observables 
    \beq
    \{ W( w, \Sigma), W(\lambda, \Sigma)\} = -W([w, \lambda]_s, \Sigma)
    \eeq{bvvbsuiw9900}
    get mapped into the Schouten bracket between the multivector fields \cite{Cattaneo:1999fm}.

\subsection{General comments on the path integral}
\label{general}

The main task is to calculate the correlation functions of observables which can be represented
 as the path integral expression 
\beq
 \langle W(w_1, c_{i_1})... W(w_n, c_{i_n})\rangle =  \int\limits_{\cal L} {\cal D}{\mathsf X}{\cal D}
  \boldsymbol{\eta} \,\,W(w_1, c_{i_1})... W(w_n, c_{i_n})\,\, e^{\frac{i}{\hbar}S_{BV}}~.
\eeq{corlalsd;3939}
  For this integral to make sense at least perturbatively  we have to integrate not over whole functional space but over the "Lagrangian" submanifold ${\cal L}$. The choice of ${\cal L}$ is called the gauge 
  fixing and it is typically generated by a gauge fixing fermion $\Psi$.
   The path integral (\ref{corlalsd;3939}) is invariant under the deformations of the Lagrangian submanifold ${\cal L}$. 
  
  However due to the absence of any well-defined measure on the space of fields we cannot treat this integral non-perturbatively. Despite this difficulty we can address and even sometimes to solve it completely 
    from the different direction, namely by reducing to an appropriate finite dimensional problem. We would expect that the correlator (\ref{corlalsd;3939}) has a well-defined 
     expansion in non-negative powers of $\hbar$.  In particular there will be a leading term in this expansion which we can evaluate by consistent reduction of the full theory to a finite dimensional BV
      theory for which all objects can be defined.  This reduction will produce the leading terms in the 
       correlators. Indeed for some models these terms correspond to a full quantum result. In the Section
        \ref{multivectors} we will consider the finite dimensional BV theory responsible for a leading 
         terms in the correlators on $S^2$.
        
  In Section \ref{gauge}  we present the details for a concrete choice of ${\cal L}$. 
   The gauge fixed theory will have residual BRST symmetry which allows us to localize the infinite 
    dimensional integrals to finite dimensional.

 \section{The reduced BV theory}
\label{multivectors}

 In this Section we consider a consistent truncation of the infinite dimensional BV theory to a finite dimensional one, that computes the contribution of constant configurations. We conjecture that this reduced BV theory controls  the leading contribution into
  the path integral in the limit $\hbar \rightarrow 0$. 
 
  This procedure can be considered as a reduction of $BV$-manifolds and
  for a Riemann surface $\Sigma_g$ of genus $g$ the truncation can be organized in the following fashion.  We define the submanifold $\cal C$ of the whole space of fields by requiring that all fields are closed forms
  \begin{equation}\label{constraints} dX=0~,~~~d\beta=0~,~~~d\eta=0~,~~~d\eta^+=0~,~~~dX^+=0~,~~~d\beta^+=0~.\end{equation}
  These  equations define a set of first class constraints (the conditions  $dX^+=d\beta^+=0$ are redundant 
   since  $X^+$ and $\beta^+$ are the top forms), i.e. $\cal C$ is a coisotropic submanifold.
    The gauge transformations generated
    by the constraints (\ref{constraints}) shift the field by  an exact form. 
     Therefore the reduced BV space is obtained by going to the cohomology of $\Sigma_g$. The reduced variables are then defined by the integration of  the fields over all cycles of $\Sigma_g$. 
     Thus zero-forms $X$ and $\beta$ are constants, and we use the same symbols to indicate the reduced coordinates. For one-forms we choose the basis $\{c_a\}$ in  $H_1(\Sigma_g, \mathbb{R}) = H^1(\Sigma_g, \mathbb{R})$ and introduce the reduced coordinates
   $$ \eta_a = \int\limits_{c_a}\eta ~,~~~~~~~\eta^+_a =  \int\limits_{c_a}\eta^+~.$$ 
   While two-forms $X^+$ and $\beta^+$ are integrated over whole $\Sigma$ and give 
       $$ X^+ =\int\limits_{\Sigma_g} X^+~,~~~~~~~~~~\beta^+ =\int\limits_{\Sigma_g} \beta^+~.$$ 
All the $BV$ structure goes to the quotient and defines a finite dimensional $BV$ manifold. The space $H^1(\Sigma_g, \mathbb{R})$ is symplectic with the structure $\omega^{ab}$. 
Therefore on the reduced finite dimensional manifold, the odd symplectic structure (\ref{oddsymplecticsty}) reads
       \beq
      \omega = d X^\mu d X^+_\mu +  \omega^{ab} d\eta_a d\eta^+_{b}      + d\beta_\mu d\beta^{+\mu}~.
     \eeq{oddbecome002888}
 Moreover, the $BV$ action $S_{BV}$ defined in (\ref{BVfullaction}) when restricted to ${\cal C}$ depends only on the reduced variables, {\it i.e.} it is a pull-back of a function on the reduced manifold. We use the same notation $S_{BV}$ for it.

 However  we are interested in zero genus case, and we leave for future investigations the case of genus $g>0$. In this situation the corresponding finite dimensional BV manifold  is ${\cal F}=T^*[-1] T^*[1]M$ where the odd symplectic structure is written in the coordinates $z=(X^\mu,\beta_\mu,X^+_\mu,\beta^{+\mu})$ as 
\beq
 \omega = d X^\mu d X^+_\mu + d\beta_\mu d\beta^{+\mu}~.
\eeq{formalllssusi329}
The degree of the coordinates is the one induced from the corresponding fields.
Under a coordinate change $\tilde{X}^i(X^\mu)$, the new coordinates $\tilde{z}=(\tilde{X}^i,\tilde\beta_i,{\tilde X}^{+i},\tilde\beta^{\dagger i})$ are 
\begin{equation}\label{changeofcoordinates}
\tilde\beta_i=T_i^\mu\beta_\mu~,~~~~ \tilde\beta^{+i}=T^i_\mu \beta^{+\mu}~,~~~~\tilde{X}^+_i = X^+_\mu T^\mu_i - \beta^{+\mu}\beta_\nu \frac{\partial T^\nu_j}{\partial Y^i}  (T^{-1})^j_\mu\;,
\end{equation}
where $T^\mu_i=\partial X^\mu /\partial \tilde X^i$.
The BV action (\ref{BVfullaction}) becomes 
 \beq
  S_{BV} =   X^+_\mu \alpha^{\mu\nu}(X) \beta_\nu  
  -\frac{1}{2} \beta^{+\mu} \partial_\mu \alpha^{\rho\nu}(X)\beta_\rho\beta_\nu  ~ ,
  \eeq{BVactionfindikksk}
 which obviously satisfies the classical master equation.  In the following discussion we will analyze this 
  finite dimensional BV theory and claim that it gives the leading contribution to PSM correlators.
   Later using a particular gauge fixing we will confirm this statement. 
   
   In addition to the BV reduction described above we can provide a different heuristic argument in 
    the support of our construction. The action (\ref{BVactionfindikksk}) can be understood
     as a leading term in the effective BV theory with the "constant" maps as IR degrees of freedom. 
      The reader may consult  \cite{Krotov:2006th, Mnev:2006ch} for the explanation 
      the effective actions within the BV framework.
        
 \subsection{Integration on finite dimensional BV manifold}

We start by defining the integration over ${\cal F}=T^*[-1] T^*[1]M$. This will allow us to define an odd Laplacian which is necessary for a proper BV description, according to the lines outlined in Section \ref{reviewBV}.

   Integration on ${\cal F}$ can be defined by putting together berezinian integration in the odd directions of $X^{+}_\mu$ and $\beta_\mu$ and fiberwise integration in the even directions of $\beta^{+\mu}$. Let us choose a volume form $\Omega=\Omega_{\mu_1\cdots \mu_n}dX^{\mu_1}\cdots dX^{\mu_n}=\rho_\Omega dX^1\cdots dX^n$ on $M$. 
  

We introduce the volume form $\mu_\Omega=\rho_\Omega^4 Dz$, where $Dz=dX^1\cdots d\beta_1\cdots dX^+_1\cdots d\beta^{+1}\cdots$ is the coordinate volume form. Since under the change of coordinates (\ref{changeofcoordinates}) the coordinate volume form transforms as
$$D\tilde{z}={\rm Ber}\frac{\partial \tilde z}{\partial z} Dz~, ~~~~ {\rm Ber}\left(\begin{array}{cc}I_{00}& I_{01}\cr I_{10}&I_{11}\end{array}\right)=\frac{\det(I_{00}-I_{01}I_{11}^{-1}I_{10})}{\det I_{11}}$$
it is simple to check that $\mu_\Omega$ is well defined. By applying (\ref{global_laplacian}), we get
$$\Delta_\Omega= \frac{\d}{\d X^+_\mu}\frac{\d}{\d X^\mu} - \frac{\d}{\d \beta^{+\mu}}\frac{\d}{\d \beta_\mu} + 2\{\log\rho_\Omega,-\}\;. $$

The restriction to ${\cal F}$ of local and the non-local observables (\ref{38383iiiskkka}) associated to multivector fields defines the corresponding observables on the reduced manifold $\cal F$. Namely, to $w\in\Gamma(\wedge^pTM)$ we associate the local observable 
\beq
O_0^p(w)=w^{\mu_1\cdots\mu_p}\beta_{\mu_1}\cdots\beta_{\mu_p}~,
\eeq{localobservable}
and the non-local one 
\beq
O^{p-2}_2(w)= -\partial_\rho w^{\mu_1\cdots\mu_p}\beta^{+\rho}\beta_{\mu_1}\cdots \beta_{\mu_p}+pw^{\mu_1\cdots\mu_p}X^+_{\mu_1}\beta_{\mu_2}\cdots\beta_{\mu_p}~.
\eeq{nondll3w990230}
It is straightforward to check that they are covariant under the transformation of coordinates (\ref{changeofcoordinates}). The antibracket defined by the odd symplectic structure (\ref{oddbecome002888}) 
  between local and non-local observables can be expressed in terms of the Schouten bracket; let $w\in\Gamma(\Lambda^pTM)$, $\lambda\in\Gamma(\Lambda^\ell TM)$,  then we have that
  \beq
\{O_2^{p-2}(w),O_0^\ell(\lambda)\}=-O_0^{p+\ell-1}([w,\lambda]_s)~~~~~~~\{ O^{p-2}_2(w), O^{\ell-2}_2(\lambda) \} = - O^{p+\ell-3}_2([w,\lambda]_s)~,
  \eeq{Pdsdkdkkwowow}
   in analogy with (\ref{bvvbsuiw9900}). 
The odd Laplacian $\Delta_\Omega$ acts on this observable as follows
\begin{equation}\label{derived_laplacian}
 \Delta_{\Omega} O^{p-2}_2(w)= -2
 (D_\Omega(w))^{\mu_1\cdots\mu_{p-1}}\beta_{\mu_1}\cdots\beta_{\mu_{p-1}}=-2O^{p-1}_0(D_\Omega(w))~,
\end{equation}
 where $D_\Omega$ is the divergence associated to the volume form $\Omega$ defined in the Appendix A. The $BV$-differential also descends to the reduced manifold as $\delta_{BV}(F)=\{S_{BV},F\}$, for any $F\in C^\infty({\cal F})$. 

The action $S_{BV}=1/2 \ O_2^0(\alpha)$ defined in (\ref{BVactionfindikksk}) satisfies the quantum master  equation (\ref{rela327388}) if the following holds 
\beq
 \Delta_\Omega S_{BV} + \frac{1}{2}\{ S_{BV}, S_{BV} \}=0~~\Longleftrightarrow~~D_\Omega \alpha =0~,~~~ [\alpha, \alpha]_s=0~,
\eeq{quanuieiowek}
 where $[~,~]_s$ is the Schouten bracket on multivector fields, see Appendix \ref{a:PL} for the definitions. 
   Thus the classical and quantum master equations have to be satisfied simultaneously. The geometrical meaning of the quantum master equation is clear: the volume form $\Omega$ must be invariant under the flow of the hamiltonian vector fields of $\alpha$. The existence of such volume form is equivalent to the unimodularity of the Poisson tensor, see the discussion in Appendix A. 
    More generally, we may say that the action (\ref{BVactionfindikksk}) is of order zero in 
     $\hbar$ of the solution of quantum master equation if and only if $\alpha$ is Poisson and 
     unimodular\footnote{Within the general BV framework it can be shown that the modular class corresponds to the first obstruction to the existence of a quantum master action \cite{Lyakhovich:2004kr}.}. If $\Omega$ is not invariant form then the unimodularity of $\alpha$ implies 
     \beq
      D_\Omega \alpha = - d_{LP} f~,
     \eeq{udneieo3993002}
     for some function $f(X)$.  This would correspond to the addition to 
     $$S_{BV} + 2 \hbar f(X)~.$$
     Equivalently this amounts to the redefinition $\Omega$ by $e^{\hbar f} \Omega$. In what 
      follows we set $\hbar=1$. 
   
By applying formulas (\ref{Pdsdkdkkwowow}), we see that for
  any $w\in \Gamma(\Lambda^\bullet TM)$ we have
  \beq
  \Delta_{(\Omega,\alpha)}O_0^p(w)=0~~~~~~\Longleftrightarrow ~~~ d_{LP}(w)=0~,
  \eeq{quantumlocalobservable}
and thus the local observable associated to $w$ is a quantum observable iff $d_{LP}w=0$. The non-local observable $O_2^{p-2}(w)$ will be quantum if the following holds
 \beq
  \Delta_{\Omega}\left  ( O^{p-2}_2(w) e^{S_{BV}}  \right )=0 ~~\Longleftrightarrow~~
   \Delta_{(\Omega,\alpha)} ( O^{p-2}_2(w))=0
   ~~\Longleftrightarrow~~
       D_\Omega w =0~,~~ d_{LP} w =0~.
\eeq{quantums002989}
Moreover, by applying (\ref{Pdsdkdkkwowow}) we see that local and nonlocal observables form a subcomplex of the quantum laplacian $\Delta_{(\Omega,\alpha)}=\Delta_\Omega+\delta_{BV}$. See 
  the next subsection for the discussion of these observables.  
   
   Finally we can evaluate the path integral. We have to choose a Lagrangian submanifold ${\cal L}$
    and the most obvious choice is ${\cal L} = \{ X^+=0, \beta^+=0\}$. In order to compensate 
     the odd integration we have to insert into the path integral the local observables 
    \beq
     \int\limits_{\cal L} O_0^{p_1} (w_1)~ .... ~O_0^{p_k} (w_k)~ e^{S_{BV}} = tr_\Omega (w_1 \wedge ... \wedge w_k)~,
    \eeq{pathinterhjjks99}
    where the trace map is defined in the Appendix \ref{a:spinors}. This expression is non-zero only
      if $p_1 +...+p_k=d$. With this choice of lagrangian submanifold, the nonlocal observables are identically zero. 

We  conclude that in the present finite dimensional $BV$-theory 
    the action (\ref{BVactionfindikksk}) satisfies the 
 quantum master equation if  the Poisson tensor $\alpha$ is unimodular.
 This is  equivalent to the requirement that there exists a trace map ${\rm tr}_\Omega$ satisfying two properties  in Theorem \ref{summarize} of Appendix A. In Appendix A we present the mathematical 
  discussion of these properties. Below we present "physical" derivation of those identities.
  The first property of $tr_\Omega$ from Theorem \ref{summarize} is a consequence of the quantum master equation for $S_{BV}$ (i.e., the unimodularity of Poisson structure $\alpha$). Namely we
   have the following chain of relations 
\begin{eqnarray*}
tr_\Omega \left (d_{LP} (w)\wedge\lambda-(-1)^{|w|+1} w\wedge d_{LP}(\lambda)\right )&=& 
tr_\Omega \left (d_{LP} (w \wedge \lambda )\right )= \cr
 = - 2\int\limits_{\cal L}  \{ e^{S_{BV}}, O_0^{|w|+|\lambda|} (w\wedge \lambda) \} &=& 
 - 2\int\limits_{\cal L} \Delta_\Omega \left (e^{S_{BV}} O_0^{|w|+|\lambda|} (w\wedge \lambda)\right )=0~.
\end{eqnarray*}
 This property implies that the trace map $tr_\Omega$ descends to the Lichnerowicz-Poisson cohomology $H^\bullet_{LP} (M)$.
The second property in Theorem \ref{summarize} is  a simple consequence of the fundamental $BV$ Theorem \ref{BVtheorem}.  To be specific for the multivector fields $w,\lambda$ we have 
 the following relations
\begin{eqnarray*}
tr_\Omega\left (D_\Omega(w)\wedge\lambda-(-1)^{|w|} w\wedge D_\Omega(\lambda)\right )&=&\cr
= \int\limits_{\cal L} \left (O_0^{|w|-1}(D_\Omega w) O_0^{|\lambda|}(\lambda)-(-1)^{|w|} O_0^{|w|}(w) O_0^{|\lambda|-1}(D_\Omega\lambda) \right ) &= &\cr
=-2\int\limits_{\cal L} \Delta_\Omega\left (O_2^{|w|-2}(w)O_0^{|\lambda|}(\lambda)-O_0^{|w|}(w)O_2^{|\lambda|-2}(\lambda)\right ) &=&0~, 
\end{eqnarray*}
where (\ref{quanuieiowek}) and (\ref{quantums002989}) have been used. This property implies 
 that the trace descends to the cohomology of $D_\Omega$.  The cohomology of $D_\Omega$ on 
  the multivectors $H^\bullet (D_\Omega)$ is isomorphic to the de Rham cohomology $H^\bullet_{dR} (M)$. 
  
  In the present context it is worthwhile to mention another interesting property of the trace
    map $tr_\Omega$ on multivector fields. For the unimodular Poisson structure $\alpha$ 
     there is the following relation 
     \beq
      e^{-\alpha} D_\Omega e^\alpha = d_{LP} + D_\Omega~,
     \eeq{defkewk992300}
      where $e^\alpha$ acts on the multivector field $w$ as
      $$ e^\alpha w = w + \alpha \wedge w + \frac{1}{2} \alpha \wedge \alpha\wedge w + ... ~,$$
       and $D_\Omega e^\alpha = 0$ is used.  The relation (\ref{defkewk992300}) implies the isomorphism 
        of cohomologies, $H^\bullet (d_{LP} + D_\Omega) \approx H^\bullet_{dR}(M)$. 
        Moreover the trace map 
         $tr_\Omega$ descends to the cohomology $H^\bullet (d_{LP} + D_\Omega)$.


     

     
     



\subsection{Maurer-Cartan equation and formal Frobenius manifolds}

In this subsection we comment on the relation between the $BV$ setting described above and  
the construction of Frobenius manifolds from $BV$-manifolds which appeared previously in mathematical works, in particular in the papers by Barannikov and Kontsevich \cite{BK} and by Manin \cite{manin, maninbook}.  Our observations have preliminary and speculative character. 
 We plan to come back to this subject elsewhere. 

The $BV$ theory discussed in the previous section can be deformed by adding to the solution (\ref{BVactionfindikksk}) of the quantum master equation any observable of ghost number 0.
  Take $w(t)\in \Gamma (\Lambda^2TM[[t]]) $ with $t$ being a formal parameter of degree zero such that  $w=w(0)$.  Consider the deformed action 
\beq
 S_{BV}(t) = S_{BV} + \frac{t}{2} O_2^{0}(w(t))~.
\eeq{newBVact1}
Obviously, $S_{BV}(t)$ satisfies the quantum master equation if and only if $\alpha+tw(t)$ is an unimodular Poisson structure with the invariant volume form $\Omega$. 
 This is equivalent to the Maurer-Cartan equation for $w(t)$,
\begin{equation}
 \label{maurer_cartan}
 d_{LP}w(t) + \frac{t}{2} [w(t),w(t)]_s = 0 ~,~~~~~~~ D_\Omega w(t)=0~.
\end{equation}
At the infinitesimal level this means $d_{LP}w= D_\Omega w=0$ and thus
  $O_2^0(w)$ is a quantum non-local observable.  However it is natural to allow 
   the volume form $\Omega$ to vary and use the argument presented
     around the equation (\ref{udneieo3993002}).  Therefore we can describe 
      the infinitesimal deformations as follows
      \beq
       d_{LP} w =0~,~~~~~~~~~~D_\Omega w + d_{LP} f =0~,
      \eeq{esjxkxo30393993}
   with $w + f \in \Gamma(\wedge^2 TM \oplus \wedge^0 TM)$, where $w$ corresponds to the deformations  of unimodular Poisson structure and $f$ to the deformations of the volume form.  
    The equations (\ref{esjxkxo30393993}) can be equivalently rewritten as follows 
    \beq
     (d_{LP} +D_\Omega ) (w+ f) = e^{-\alpha} D_\Omega e^\alpha (w+ f) = 0~,
    \eeq{eueui3999020mxll}
     where we assume that $\Omega$ is invariant volume form for $\alpha$. 
      In BV theory the deformation will be trivial if it is in the image
        of the quantum Laplacian $\Delta_{(\Omega, \alpha)}$. 
      However the question is to understand  the geometrical description of these trivial BV deformations.  
        For example, the diffeomorphisms give a trivial deformation of the $BV$ theory. 
    Namely for $w={\cal L}_\xi \alpha = d_{LP}(\xi)$ and $f=D_\Omega \xi$ for  $\xi\in \Gamma(TM)$ 
     the deformation  is trivial,  
   $$ \frac{1}{2} O_2^0(w) +  2 O_0^0 (f) = -\Delta_{(\Omega,\alpha)}O_2^{-1}(\xi)~.$$
    However the formula (\ref{eueui3999020mxll})  suggests that  
     the deformations is trivial if 
    \beq
     w + f = (d_{LP} + D_\Omega) \xi = e^{-\alpha} D_\Omega (e^\alpha \xi) ~,
    \eeq{trivial29291}
     with $\xi \in \Gamma(\wedge^\bullet TM)$, not just simply a vector.  One has to show that 
      the corresponding deformations of the BV theory are trivial. Unfortunately we are unable to 
       do it in all generality. Nevertheless we give some plausible arguments in its favor and 
        analyze the problem in special cases. 
    
   The linear space of deformations defined as the condition (\ref{eueui3999020mxll}) modulo
    the identification (\ref{trivial29291}) would be interpreted as the tangent space to some kind of 
     modular space of unimodular Poisson structures (if such space exists). 
     The crucial point motivated by the $BV$ consideration  is  that the Poisson tensors may be
       equivalent even if they are not diffeomorphic. Indeed the
         equivalence relation (\ref{trivial29291}) looks  very natural in terms of 
          the pure spinor description (see Appendix B for the details). The unimodular Poisson structure
           can be described in terms of closed pure spinor $\rho = e^\alpha \Omega$. The deformation 
            of the pure spinor would be given by 
            $$ \delta \rho = (w + f) \cdot \rho~,$$ 
            where the finite deformation is $e^{\alpha + w} e^f \Omega$. 
       The property (\ref{eueui3999020mxll}) implies that $d (\delta \rho )=0$.
         If the deformation satisfies (\ref{trivial29291}) then 
       $$ \delta \rho = (w + f) \cdot \rho = - d \left ( \xi \cdot \rho \right )~,$$
       where we used the Theorem \ref{theorem13} in the Appendix B. 
    Thus we look at the deformations of closed pure spinor modulo exact ones which correspond 
     to the subspace of the de Rham cohomology group, namely 
     $$\{ [(w+f)\cdot \rho] \in H^\bullet_{dR}(M)~,~~(w+f)\in \Gamma(\wedge^2 
     TM \oplus \wedge^0 TM)\}~,$$
      where we deal the alternative grading of the differential forms, see Appendix B. 
     Following a standard terminology, we refer to  the corresponding  space of 
      deformations of the $BV$ theory modulo the trivial ones as the {\it geometric moduli space}.
      
     Let us get back to the BV theory.  More generally we want to understand 
       the subspace of the cohomology of the quantum Laplacian spanned by non-local observables
 $$H_{nonloc}(\Delta_{(\Omega,\alpha)}) =\{[O_2(w)]\in H(\Delta_{(\Omega,\alpha)}), w\in\Gamma(\Lambda^\bullet TM)\}~.$$
 In particular we want to understand if it is finite dimensional and moreover related to the 
  de Rham cohomology  $H_{dR} (M) \approx H(D_\Omega) \approx H(d_{LP} + D_\Omega)$. 
    We are unable to answer this question in all generality. However we can analyze two 
     special cases which give a positive answer.  
  
   Let us discuss first the case of the trivial Poisson structure, $\alpha=0$. In this case a quantum non-local observable $O^{p-2}_2 (w)$ corresponds to the 
    multivector field $w\in \Gamma(\wedge^p TM)$ such that  $D_\Omega w=0$. 
   Then we can show that $O_2^{p-2}(D_\Omega\nu)$, $D_\Omega\nu\in\Lambda^pTM$, is trivial. In fact it is always possible to write $\nu=\sum_i f_i D_\Omega \lambda_i$, for some $f_i\in C^\infty(M)$ and $\lambda_i\in\Gamma(\Lambda^{p+2} TM)$. This is obviously equivalent to say that the de Rham differential finitely generates the module of forms. Then using the basic properties of the antibracket 
    we arrive to
\begin{eqnarray}
O_2^{p-2}(D_\Omega\nu) &=& \sum_i O_2^{p-2}([f_i,D_\Omega\lambda_i]_s) = -\sum_i\{O_2^{-2}(f_i),O_2^{p-1}(D_\Omega\lambda_i)\}\cr  
             &=& -\Delta_\Omega(\sum_i O_2^{-2}(f_i)O_2^{p-1}(D_\Omega\lambda_i))~.
\end{eqnarray}
 Therefore  the correspondence $w\rightarrow O_2^{p-2}(w)$ defines a surjection from $H(D_\Omega)$ to  $H_{nonloc}(\Delta_{\Omega})$.  Thus the corresponding geometrical 
  moduli space is finite dimensional. 
 
 Next consider the case of non-trivial Poisson structure $\alpha$ such that
  two differentials $(d_{LP},D_\Omega)$ 
 satisfy the $\partial\bar\partial$-lemma, {\it i.e.}
\begin{equation}\label{lemmaddbar}
 {\rm Im} d_{LP} D_\Omega = {\rm Im} d_{LP}\cap{\rm Ker} D_\Omega = {\rm Ker} d_{LP}\cap{\rm Im} D_\Omega\;.
\end{equation}
The condition (\ref{lemmaddbar}) is satisfied for a large class of symplectic manifolds obeying the  strong Lefschetz property (see \cite{maninbook}). 
 However the $\partial\bar\partial$-lemma does not hold for a generic Poisson manifold
  since $H_{LP} (M)$ is infinite dimensional.  One  of the consequence of 
   the $\partial\bar\partial$-lemma is the isomorphism of the cohomologies, $H_{LP}(M) \approx H_{dR}(M)$.  The extreme example of the failure for this lemma is the trivial Poisson structure. 
   Consider $w\in\Gamma(\Lambda^pTM)$ which defines a trivial class in $(d_{LP}+D_\Omega)$-chomology, {\it i.e.}  $w=d_{LP}\xi_{p-1}+D_\Omega \xi_{p+1}$, $0=d_{LP}\xi_{k-1}+D_\Omega\xi_{k+1}$ for $k\not=p$. After the straightforward calculation  we arrive to the following relation
$$
O_2^{p-2}(w)=  -2 \Delta_{(\Omega,\alpha)}(O_2(\xi_{p-1})+ 4 O_0(\xi_{p-3})) + O_2(D_\Omega\xi_{p+1}) ~.
$$ 
Since $D_\Omega\xi_{p+1}\in {\rm Im}D_\Omega\cap {\rm Ker}d_{LP}={\rm Im} D_\Omega d_{LP}$  there exists $\nu_p$ such that $D_\Omega\xi_{p+1}=D_\Omega d_{LP}\nu_p$ and $O_2(D_\Omega\xi_{p+1})= 2 \Delta_{(\Omega,\alpha)}O_2(D_\Omega\nu_p)$.  Thus we 
  conclude that also in this case the correspondence
$w\rightarrow O^{p-2}_2(w)$ defines a surjective map from the finite dimensional space
  $H^p_{dR}(M, \alpha)$ to  $H_{nonloc}(\Delta_{(\Omega,\alpha)})$ where $H^p_{dR}(M)$ is 
  defined as follows
  $$ H^p_{dR}(M, \alpha) =\{ [ w\cdot \rho] \in H^\bullet_{dR}(M)~,~~ w\in \Gamma(\wedge^p TM)\}~.$$

Motivated by these two examples we conjecture that 
 the space  $H_{nonloc}(\Delta_{(\Omega,\alpha)})$ is finite dimensional.
  Thus in general the action $S_{BV}$ can be deformed for arbitrary ghost number, mimicking of the construction of Frobenius manifolds of \cite{BK} and \cite{manin}. Let $\{w_k\in \Gamma(\Lambda^{p_k} TM)\}$ define a basis $\{O_2^{p_k-2}(w_k)\}$ of $H_{nonloc}(\Delta_{(\Omega,\alpha)})$. We  introduce the formal variables $\{t_k\}$ of degree $2-p_k$ and extend the full $BV$ machinery to ${\cal F}\otimes {\mathbb R}[[t_k]]$. Clearly $S(t)=S_{BV}+\sum_k t_k O_2^{p_k-2}(w_k)$ solves at the infinitesimal level the quantum master equation.  Interpreting  $H_{nonloc}(\Delta_{(\Omega,\alpha)})$ as the tangent space of the  {\it extended moduli space} the main problem is to find a finite deformation, {\it i.e.} a solution of the Maurer-Cartan equation
\begin{equation}\label{extended_maurer_cartan}
 \delta_{BV} S(t) + \frac{1}{2} \{S(t),S(t)\}=0 ~.
\end{equation}
 In \cite{BK, manin, maninbook} the solution of such equation is discussed within the BV setup.
The main difference with the setup in \cite{BK, manin, maninbook} is the requirement of $\partial\bar\partial$-lemma that we want to avoid because it excludes the non symplectic cases. Is it possible to solve the Maurer-Cartan equation (\ref{extended_maurer_cartan}) in this context ? The $\partial\bar\partial$-lemma provides the isomorphism between the spaces of the  classical and quantum observables.  While for the generic unimodular Poisson manifold the space of classical observables 
 is infinite dimensional and the space of quantum observables  is expected to be finite dimensional. 


\section{Gauge fixing}
\label{gauge}

In this Section we perform the gauge fixing by choosing an appropriate Lagrangian submanifold. 
 In particular we use a complex structure for the gauge fixing.

\subsection{Geometrical setup}
\label{geometryset}

Let us start from the description of the relevant geometric setup. It turns out to be very 
 convenient to consider the $N=2$ supersymmetric PSM \cite{Calvo:2005ww}. 
  The existence of the extended supersymmetry for PSM requires a generalized complex
   strucrure
   \beq
{\cal J} = \left (\begin{array}{cc}
J & P \\
L & - J^t \end{array} \right )~,
\eeq{defdcomsp200}
 such that  $[ {\cal R}, {\cal J}] =0$, where 
\beq
 {\cal R} = \left (\begin{array}{cc}
  1_d & \alpha\\
  0 & - 1_d \end{array}\right )~.
\eeq{definibialreald}
 These conditions can be worked out completely. To be specific 
 $L=0$, $J$ is a complex structure and moreover the $(2,0)+(0,2)$ part of $\alpha$ 
 \beq
  P = \frac{1}{2} (J\alpha + \alpha J^t)~,
 \eeq{defkklal39390}
 is a holomorphic Poisson structure.  If we switch to the complex 
  coordinates with the labels $(i, \bar{i})$ then $(2,0)$-part $\alpha^{ij}$ is a holomorphic 
   Poisson structure if the following holds
   \beq
    \d_{\bar{k}} \alpha^{ij}=0~,~~~~~~~\alpha^{il}\d_l\alpha^{jk} + \alpha^{jl}\d_l\alpha^{ki}+\alpha^{kl}\d_l\alpha^{ij}=0~.
   \eeq{definaslkopo}
    Indeed the geometrical setup we will use  can be summarized as follows: a Poisson manifold $(M, \alpha, J)$ with a  complex structure $J$ such that $(2,0)$-part of $\alpha$ is holomorphic.  The fact that $(2,0)$-part 
     is Poisson itself follows from this. 

 It may look at first that the geometry we just described is somewhat exotic. However
  it is not the case  and this Poisson geometry is realized always on (twisted) generalized K\"ahler manifolds \cite{Lyakhovich:2002kc, gualtieri1, Hitchin:2005cv}. The (twisted) generalized K\"ahler manifold can be characterized as a bihermitian 
   geometry $(g, J_+, J_-)$ where $J_\pm$ are two complex structures and $g$ is a metric which is hermitian with respect to both complex structure.   In addition there are certain integrability conditions 
    on two-forms $gJ_\pm$.  The (twisted) generalized K\"ahler manifold has 
     two real Poisson structures $\pi_\pm = (J_+ \pm J_-)g^{-1}$ \cite{Lyakhovich:2002kc}. 
      Moreover their $(2,0)$-part with respect to $J_+$ (or $J_-$) is a holomorphic Poisson structure
       with respect to $J_+$ (or $J_-$),  \cite{Hitchin:2005cv}.

\subsection{Gauge fixed action}

  Let us assume that the Poisson manifold $(M, \alpha)$ admits a complex structure $J$
   such that $(2,0)$-part of $\alpha$ is a holomorphic Poisson structure and the world-sheet $\Sigma$
    is equipped with   a complex  structure.  We will concentrate our attention on the case of the two-sphere
     where the complex structure is unique. 
 Introducing the complex coordinates on $M$ and $\Sigma$ we define 
  the following Lagrangian submanifold in the space of (anti)fields
\beq
 \eta_{zi}=0,\,\,\,\,\,\,\,\eta_{\bar{z}\bar{i}}=0,\,\,\,\,\,\,\,\eta_{z}^{+i}=0~,\,\,\,\,\,\,\,
 \eta_{\bar{z}}^{+\bar{i}}=0,\,\,\,\,\,\,\,X^+=0,\,\,\,\,\,\,\,\beta^+=0~,
\eeq{defLaghwkk}
  where $(i, \bar{i})$ stand for the complex coordinates on $M$ and $(z,\bar{z})$ are the complex 
   coordinates on $\Sigma$.  The odd symplectic structure
     (\ref{oddsymplecticsty}) is zero on (\ref{defLaghwkk}). 
    Equivalently we could write the conditions (\ref{defLaghwkk}) using the projectors constructed 
     out of $J$ and complex structure on $\Sigma$, in the same fashion as in \cite{Witten:1988xj}. 
      Indeed we do not need to assume that $J$ is integrable, it is enough for $J$ to be an almost complex structure. However in what follows we are in the geometrical setup described in the previous subsection.
       In this case many calculations simplify drastically.
   
   Assuming the gauge (\ref{defLaghwkk})  the gauge fixed action is
 $$ S_{GF} = i \int\limits_{\Sigma} d^2\sigma\,\,\left [ \eta_{z\bar{i}} \d_{\bar{z}} X^{\bar{i}} - \eta_{\bar{z}i}
  \d_{z} X^i + \alpha^{\bar{i}j}\eta_{z\bar{i}} \eta_{\bar{z}j} + \eta^{+i}_{\bar{z}} (\d_z \beta_i
   + \d_i \alpha^{\bar{l}s} \eta_{z\bar{l}}\beta_s) - \right .$$
   \beq
 \left .  - \eta^{+\bar{i}}_z(\d_{\bar{z}} \beta_{\bar{i}} +
    \d_{\bar{i}}\alpha^{l\bar{s}} \eta_{\bar{z}l}\beta_{\bar{s}}) - \d_{\bar{i}} \d_j \alpha^{k\bar{l}}
    \eta^{+\bar{i}}_z \eta^{+j}_{\bar{z}} \beta_k \beta_{\bar{l}} \right ]~,
 \eeq{gaugefixedjskl}
  which is just the action (\ref{BVfullaction}) restricted to (\ref{defLaghwkk}).
  The action (\ref{gaugefixedjskl}) is invariant under the following 
BRST transformations 
\ber
\label{BB1}&& \delta X^i = \alpha^{ij}\beta_j +\alpha^{i\bar{j}}\beta_{\bar{j}}~,\\
&&\delta X^{\bar{i}} = \alpha^{\bar{i}\bar{j}}\beta_{\bar{j}} + \alpha^{\bar{i}j}\beta_j~,\\
&& \delta \eta_{\bar{z}}^{+i} = -\d_{\bar{z}} X^i - \alpha^{ij} \eta_{\bar{z}j} - \d_k \alpha^{i\bar{j}} \eta_{\bar{z}}^{+k} \beta_{\bar{j}} - \d_k \alpha^{ij} \eta_{\bar{z}}^{+k}\beta_j~,\\
&&\delta \eta_z^{+\bar{i}} = - \d_z X^{\bar{i}} - \alpha^{\bar{i}\bar{j}} \eta_{z\bar{j}} -
 \d_{\bar{k}} \alpha^{\bar{i}j} \eta^{+\bar{k}}_z \beta_j -  \d_{\bar{k}}\alpha^{\bar{i}\bar{j}}
 \eta_z^{+\bar{k}}\beta_{\bar{j}}~,\\
&& \delta \beta_i = \d_i \alpha^{k\bar{j}}\beta_k \beta_{\bar{j}} + \frac{1}{2} \d_i\alpha^{kj}\beta_k \beta_j~,\\
&& \delta \beta_{\bar{i}} = \d_{\bar{i}} \alpha^{\bar{k}j}\beta_{\bar{k}}\beta_j + \frac{1}{2}\d_{\bar{i}}\alpha^{\bar{k}\bar{j}}\beta_{\bar{k}}\beta_{\bar{j}}~,\\
\nonumber&&\delta \eta_{z\bar{i}}= -\d_z \beta_{\bar{i}} -\d_{\bar{i}} \alpha^{\bar{k}l} \eta_{z\bar{k}}\beta_l -
\d_{\bar{i}}\alpha^{\bar{k}\bar{l}}\eta_{z\bar{k}}\beta_{\bar{l}} - \d_{\bar{i}}\d_{\bar{s}} \alpha^{k\bar{l}}
\eta_z^{+\bar{s}}\beta_k \beta_{\bar{l}} -\\
&&\,\,\,\,\,\,\,\,\,\,\,\,\,\,\,\,\,\,- \frac{1}{2} \d_{\bar{i}}\d_{\bar{s}} \alpha^{\bar{k}\bar{l}}\eta_z^{+\bar{s}}\beta_{\bar{k}} \beta_{\bar{l}}~,\\
\nonumber&&\delta \eta_{\bar{z}i}= -\d_{\bar{z}} \beta_{i} -\d_{i} \alpha^{k\bar{l}} \eta_{\bar{z}k}\beta_{\bar{l}} -
\d_{i}\alpha^{kl}\eta_{\bar{z}k}\beta_{l} - \d_{i}\d_{s} \alpha^{\bar{k}l}
\eta_{\bar{z}}^{+s}\beta_{\bar{k}} \beta_{l} -\\
\label{BB7}&&\,\,\,\,\,\,\,\,\,\,\,\,\,\,\,\,\,\,- \frac{1}{2} \d_{i}\d_{s} \alpha^{kl}\eta_{\bar{z}}^{+s}\beta_{k} \beta_{l}~,
\eer{equak2929200}
 which are nilpotent only on-shell. The existence of such residual 
 BRST symmetry within BV formalism is discussed in \cite{henneaux, Alexandrov:1995kv}. 

 Next using the gauge fixed action (\ref{gaugefixedjskl}) we can calculate the path
   integral explicitly on the sphere. 
    In particular let us perform the one-loop calculation around the constant map.
    We take a classical solution $\eta=0$ and $X=x_0$ with $x_0$ being a constant 
     and the rest of fields are zero. Consider the fluctuations around this configuration 
  \beq
   X= x_0 + X_f~,~~~\eta = 0 + \eta_f ~,~~~\beta=0 + \beta_f~,~~~\eta^+ = 0 + \eta^+_f ~,
   \eeq{flucjsoowopppq}
 where naturally by $\eta$ and $\eta^+$ we understand only non-vanishing components 
  $(\eta_{\bar{z}i},\eta_{z\bar{i}})$ and $(\eta_{\bar{z}}^{+i}, \eta^{+\bar{i}}_z)$ correspondently. 
  We take the expansion (\ref{flucjsoowopppq}) and plug it into the gauge fixed 
   action (\ref{gaugefixedjskl})
   while keeping only up to the quadratic terms in the fluctuations. The bosonic part of 
    resulting action can be written schematically as 
    \beq
     \frac{1}{2} \left (\begin{array}{cc}
     X & \eta 
     \end{array}
     \right ) \left ( \begin{array}{cc}
     0 & D\\
     -D  & A
     \end{array} \right ) 
     \left ( \begin{array}{c}
     X\\
     \eta
     \end{array} \right )~,
    \eeq{bosoeie09300}
     where $A$ is a part composed from the Poisson tensor $\alpha$ and $D$ is a first 
      order differential operator
      $$ D= \left ( \begin{array}{cc}
      \d_z & 0\\
      0 & -\d_z
      \end{array} \right )$$
    While the fermionic part of the corresponding action is written as
    \beq
     \eta^t D \beta~,
    \eeq{ferioos9903034e71}
 with the same $D$.  We can perform easily the gaussian integral over the
  bosonic (\ref{bosoeie09300}) and the fermionic parts  (\ref{ferioos9903034e71}).
   The integration produces the ratio of determinants of $D$ which is exactly 1. 
    Thus the result of this gaussian integration is just one. However the integration over zero
      modes of $D$  will remain. The fields $\eta$ and $\eta^+$ do not have any zero modes 
       since there are no (anti)holomorphic 1-forms on the sphere. While $\beta$ have constant 
        zero modes and $X$ does as well. These zero modes give an integration over the finite 
         dimensional graded manifold $T^*[1]M$ which is defined by choosing a 
          volume form $\Omega$ on $M$. In order to compensate the odd integration we have 
           to insert the local observables into the path integral. Thus the final result for the correlators
            of local observales is 
            \beq
             \langle O_0^{p_1} (w_1)~ .... ~O_0^{p_k} (w_k)\rangle=
               tr_\Omega (w_1 \wedge ... \wedge w_k)~,
               \eeq{rewiew993030202020}
    where the trace map $tr_\Omega$ is defined in the Appendix and the correlator agrees with 
     (\ref{pathinterhjjks99}).
    Since the number of zero modes
     for $\beta$ corresponds to the dimensionality of $M$ we have that the
       correlator (\ref{rewiew993030202020}) is 
      non-zero only if $p_1 +...p_k =  d$.  Moreover if we require that the correlator is invariant 
       under the BRST symmetry (\ref{BB1})-(\ref{BB7})  then the Poisson tensor $\alpha$ should be unimodular and $\Omega$ is the corresponding invariant volume form. To prove this 
       we need to remember how BRST   symmetry (\ref{BB1})-(\ref{BB7}) acts on
        the  local observables and the  theorem \ref{theoremPLinterga} from 
        the Appendix \ref{a:PL}.  Notice that as far as the fields $X$ and $\beta$ concern the action of BV symmetry (\ref{AAAA1})-(\ref{AAAA7}) and the BRST symmetry (\ref{BB1})-(\ref{BB7}) 
        is the same. Since the local observables are constructed from $X$ and $\beta$
         only we can apply the discussion of the subsection \ref{observables} to 
         the analysis of BRST invariant observables in the present setup.  
          
  We conclude that the present calculation is in complete agreement with our previous analysis 
   within the finite dimensional BV framework.  Although the unimodularity of $\alpha$ is argued 
    completely differently, now through the BRST invariance of the zero-mode measure.  
     The answer (\ref{rewiew993030202020}) is just the leading contribution into
       the full quantum correlator.     
    
    Finally we comment when the geometry required for the present gauge fixing is compatible 
     with the unimodularity.  Indeed for a generalized Calabi-Yau manifold  the corresponding Poisson 
      structure is always unimodular  \cite{gualtieri2}.  Thus as a possible example, we may consider 
       the generalized K\"ahler geometry where one of the generalized complex structures satisfies
        a generalized Calabi-Yau condition.  Actually the gauge fixing can be performed for a
         generalized Calabi-Yau manifold by itself with the use of an almost generalized 
          complex structure.  However we have to stress that unimodularity of Poisson structure is a
           real condition and indeed much weaker than the generalized Calabi-Yau condition.

\subsection{Relation to A-model}
\label{A-model}

If we assume that $\alpha^{ij}=0$ and $\alpha$ is invertible then we are on K\"ahler
  manifold where  $\omega=\alpha^{-1}$ is K\"ahler form and $g = -\omega J$ is hermitian metric.
Due to the fact that $\alpha$ is invertible we can perform the integration over
  $\eta_{z\bar{i}}$ and $\eta_{\bar{z}i}$ in in the path integral with the gauge
    fixed action (\ref{gaugefixedjskl}).
  Introducing the following notation
  \beq
   \psi^i = - i g^{i\bar{j}} \beta_{\bar{j}}~,\,\,\,\,\,\,\,\,\,\,\,\,\,\,
   \psi^{\bar{i}} = i g^{\bar{i}j} \beta_j~,\,\,\,\,\,\,\,\,\,\,\,\,\,\,
   \psi_{\bar{z}}^i = -i\eta_{\bar{z}}^{+i}~,\,\,\,\,\,\,\,\,\,\,\,\,\,\,
   \psi_z^{\bar{i}}= - i \eta_z^{+\bar{i}}
  \eeq{newnotfieodla}
  the result of the integration of $\eta$ is 
 \beq
  S_A = \int d^2\sigma\left [ \d_{\bar{z}} X^{\bar{i}} g_{\bar{i}j} \d_z X^j + i \psi_z^{\bar{i}} g_{\bar{i}j}\nabla_{\bar{z}} \psi^j + i \psi_{\bar{z}}^i g_{i\bar{j}} \nabla_z \psi^{\bar{k}} - {\cal R}_{p\bar{i}j\bar{n}}
  \psi_{\bar{z}}^j \psi_z^{\bar{i}} \psi^p \psi^{\bar{n}}\right ]~,
 \eeq{Amodelactioan}
 where we adopted  the following notation
 \beq
 \nabla_{\bar{z}} \psi^k = \d_{\bar{z}} \psi^k + \Gamma^k_{\,\,\,nl}\d_{\bar{z}}X^n \psi^l~,\,\,\,\,\,\,\,\,\,\,\,\,\,\,\,\,\,\,\,\,\,\,
 \nabla_z\psi^{\bar{k}} = \d_z \psi^{\bar{k}} + \Gamma^{\bar{k}}_{\,\,\,\bar{n}\bar{l}} \d_z X^{\bar{n}} \psi^{\bar{l}}
 \eeq{defcovarderuyq}
  with $\Gamma$ being the Levi-Civita connection and ${\cal R}$ the corresponding Riemann tensor. 
   The first term in the action (\ref{Amodelactioan}) can be rewritten as
   \beq
    \d_{\bar{z}} X^{\bar{i}} g_{\bar{i}j} \d_z X^j = \frac{1}{2} \sqrt{h} h^{\alpha\beta} \d_\alpha X^{\bar{i}}
     g_{\bar{i}j} \d_\beta X^j + \frac{1}{2} \epsilon^{\alpha\beta} \d_\alpha X^{\bar{i}} (ig_{\bar{i}j}) \d_\beta X^j~,
   \eeq{termtopkinetic}
    where the last term is a topological, the pull-back of the K\"ahler form $\omega$. 
 The  BRST transformations (\ref{BB1})-(\ref{BB7}) become
    \ber
&& \delta X^i = \psi^i~,~~~\delta X^{\bar{i}}=\psi^{\bar{i}}~,~~~ 
 \delta \psi^i =  0~,~~~\delta \psi^{\bar{i}} =  0~,\\
&&
\delta \psi_{\bar{z}}^{+i} = i \d_{\bar{z}} X^i + \Gamma^i_{\,\,\,lk} \psi_{\bar{z}}^k \psi^l~, ~~~
\delta \psi_z^{+\bar{i}} = i \d_z X^{\bar{i}} + \Gamma^{\bar{i}}_{\,\,\,\bar{l}\bar{k}}\psi_{z}^{\bar{k}}\psi^{\bar{l}} ~.
\eer{equak29blavflal}
 The action (\ref{Amodelactioan}) with the BRST transformations (\ref{equak29blavflal}) corresponds
 to the topological sigma model \cite{Witten:1988xj} on K\"ahler manifold 
    which corresponds to A-twist of  $N=(2,2)$ supersymmetric sigma model \cite{Witten:1991zz}.
     Previously the BV treatment of A-model has been discussed in \cite{Alexandrov:1995kv}.
      Here we presented the improved analysis  of the relation between the BV-formulation of 
       PSM and the A-model.
     
     Any symplectic manifold with symplectic structure $\omega$ is unimodular with the volume form 
      given by $\Omega = \omega^{d/2}$.  Moreover there exists a natural isomorphism between
       the Lichnerowicz-Poisson cohomology and the de Rham cohomology, 
     $H_{LP}^\bullet(M) \approx H_{dR}(M)$ which is provided by the symplectic structure $\omega$. 
      Therefore the observable corresponding to a multivector field can be mapped into the observable 
       corresponding to the differential form through the identification (\ref{newnotfieodla}). Thus the 
        correlator (\ref{rewiew993030202020}) can be rewritten as
        \beq
          tr_\Omega (w_1 \wedge ... \wedge w_k) = \int\limits_M (\sharp w_1) \wedge ... (\sharp w_k)~,
        \eeq{regwyuuiskkll3200} 
     where $\sharp w_l$ is a differential form corresponding to a multivector field $w_l$ constructed 
      through the map $\sharp : \wedge^\bullet TM \rightarrow \wedge^\bullet T^*M$ defined by the symplectic structure $\omega$.  Indeed the correlator (\ref{regwyuuiskkll3200}) is the standard 
       one for the A-model and can be interpreted as the intersection number of the Poincar\'e dual 
        cycles to $\sharp w_l$. In the full quantum theory the correlator (\ref{regwyuuiskkll3200}) gets corrections from the holomorphic maps on which the theory is localized. These instanton
          corrections are related to the Gromov-Witten invariants. 
         This is well-developed subject, see \cite{Hori:2003ic} for a review. 
      
 \subsection{Zero Poisson structure}
 \label{zero}
 
 As a next example we consider the case of zero Poisson structure, $\alpha=0$. In this case
 the gauge fixed action (\ref{gaugefixedjskl}) is of the form 
 \beq
  S_{GF} = i \int\limits_{\Sigma} d^2\sigma\,\,\left [ \eta_{z\bar{i}} \d_{\bar{z}} X^{\bar{i}} - \eta_{\bar{z}i}
  \d_{z} X^i + \eta^{+i}_{\bar{z}} \d_z \beta_i -  \eta^{+\bar{i}}_z \d_{\bar{z}} \beta_{\bar{i}}   \right ]~,
 \eeq{gaugefixedjsklextra}
  while the BRST transformations (\ref{BB1})-(\ref{BB7}) become
   \ber
\label{RR1}&& \delta X^i = 0~,~~~\delta X^{\bar{i}}=0~,~~~ 
\delta \eta_{\bar{z}}^{+i} = -\d_{\bar{z}} X^i~, ~~~
\delta \eta_z^{+\bar{i}} = - \d_z X^{\bar{i}} ~,\\
\label{RR2}&& \delta \beta_i =  0~,~~~\delta \beta_{\bar{i}} =  0~,~~~\delta \eta_{z\bar{i}}= -\d_z \beta_{\bar{i}} ~,~~~
 \delta \eta_{\bar{z}i}= -\d_{\bar{z}} \beta_{i} ~.
\eer{equak2929200extranew}
 Now these transformations are nilpotent off-shell.  The action (\ref{gaugefixedjsklextra})
  is reminiscent of the action obtained through the  infinite volume limit  of the A-model 
  \cite{Frenkel:2005ku}.  However our BRST symmetry differs from the one
    discussed in  \cite{Frenkel:2005ku} and thus these are different theories.  
    As well the action (\ref{gaugefixedjsklextra}) 
    with the symmetries (\ref{RR1})-(\ref{RR2}) has appeared in the different context 
   in \cite{Zucchini:2005cq} as a specific gauge fixed version of "Hitchin sigma
     model" \cite{Zucchini:2004ta}. 
 
  Next we argue that the correlator (\ref{rewiew993030202020}) is a full quantum answer for 
   the PSM with $\alpha=0$.  We can use the BRST symmetry (\ref{RR1})-(\ref{RR2}) to localize 
    the theory on the holomorphic maps, $\d_{\bar{z}} X^i=0$. Namely we can add to the action
    (\ref{gaugefixedjsklextra}) the BRST exact term
    \beq
   - t  \delta \int\limits_\Sigma d^2\sigma \left ( \eta_z^{+\bar{i}} g_{\bar{i}j} \d_{\bar{z}} X^j +
    \eta_{\bar{z}}^{+i}   g_{i\bar{j}} \d_z X^{\bar{j}} \right ) =   t   \int\limits_\Sigma d^2\sigma \left (
    \d_z X^{\bar{i}} g_{\bar{i} j}\d_{\bar{z}} X^j + \d_{\bar{z}} X^i g_{i\bar{j}} \d_z X^{\bar{j}} \right )~,
    \eeq{BRSTeaxtvsjjk}
    where $t$ is any real number and this exact term is positive definite.  The addition of this exact term 
     to the action cannot change the theory and the result is independent from the parameter $t$. 
      By sending $t$ to the infinity the dominant contribution to the path integral will come from 
       the holomorphic maps, $\d_{\bar{z}} X^i=0$ and $\d_{z} X^{\bar{i}}=0$. Moreover we can perform 
        the integration over $\eta$ which impose the conditions $ \d_{\bar{z}} X^{\bar{i}} =0$ and
         $ \d_{z} X^{i} =0$ which together with the BRST argument imply that only the constant 
          maps contribute to the path integrals. Thus in the evaluation of the path integral on the sphere
            with the insertion 
           of local observables  the only remaining integration is
            the integration over $M$ and the corresponding zero modes of $\beta$. On the sphere 
             there will be no zero modes for $\eta$ and $\eta^+$.  
             
  Thus we have proven that for the PSM with zero Poisson structure the
    leading result (\ref{rewiew993030202020}) for 
   the correlators of local observables is indeed exact.  Actually this should not be surprise 
    since the Poisson tensor controls $\hbar$-corrections. In the general action (\ref{fullBVsuperfiel})
     the fields can be  rescaled in such way that $\hbar$ appears in front of $\alpha$ only. 
 
 \subsection{Holomorphic Poisson structure}
 \label{holomorphic}
 
 Another interesting case is when there exists such a complex structure $J$ that $\alpha$ is 
  a holomorphic Poisson structure. In other words $(1,1)$-part of $\alpha$ vanishes and thus
   the gauge fixed action (\ref{gaugefixedjskl}) is independent of $\alpha$. 
    The gauge fixed action for the holomorphic 
    Poisson structure is the same as (\ref{gaugefixedjsklextra}) for the zero Poisson structure
  However the Poisson structure enters into the BRST transformations. 
  For the case of holomorphic Poisson structure the transformations (\ref{BB1})-(\ref{BB7}) become 
 \ber
\label{HH1}&& \delta X^i = \alpha^{ij}\beta_j ~,\\
&&\delta X^{\bar{i}} = \alpha^{\bar{i}\bar{j}}\beta_{\bar{j}} ~,\\
&& \delta \eta_{\bar{z}}^{+i} = -\d_{\bar{z}} X^i - \alpha^{ij} \eta_{\bar{z}j} - \d_k \alpha^{ij} \eta_{\bar{z}}^{+k}\beta_j~,\\
&&\delta \eta_z^{+\bar{i}} = - \d_z X^{\bar{i}} - \alpha^{\bar{i}\bar{j}} \eta_{z\bar{j}} -  \d_{\bar{k}}\alpha^{\bar{i}\bar{j}}
 \eta_z^{+\bar{k}}\beta_{\bar{j}}~,\\
&& \delta \beta_i =  \frac{1}{2} \d_i\alpha^{kj}\beta_k \beta_j~,\\
&& \delta \beta_{\bar{i}} =  \frac{1}{2}\d_{\bar{i}}\alpha^{\bar{k}\bar{j}}\beta_{\bar{k}}\beta_{\bar{j}}~,\\
&&\delta \eta_{z\bar{i}}= -\d_z \beta_{\bar{i}} - 
\d_{\bar{i}}\alpha^{\bar{k}\bar{l}}\eta_{z\bar{k}}\beta_{\bar{l}}  - \frac{1}{2} \d_{\bar{i}}\d_{\bar{s}} \alpha^{\bar{k}\bar{l}}\eta_z^{+\bar{s}}\beta_{\bar{k}} \beta_{\bar{l}}~,\\
\label{HH7}&&\delta \eta_{\bar{z}i}= -\d_{\bar{z}} \beta_{i} - 
\d_{i}\alpha^{kl}\eta_{\bar{z}k}\beta_{l} -   \frac{1}{2} \d_{i}\d_{s} \alpha^{kl}\eta_{\bar{z}}^{+s}\beta_{k} \beta_{l}~.
\eer{equak2929200extra}
 These transformations are nilpotent 
 $\delta^2=0$ {\it off-shell} and the action (\ref{gaugefixedjsklextra}) is invariant under them. 
  Indeed there is not single BRST transformation but a whole family. In the transformations
    (\ref{HH1})-(\ref{HH7}) we can put a complex parameter $t \in \mathbb{C}$ in front of all terms containing $\alpha^{ij}$ and correspondently $\bar{t}$ in front of terms with  $\alpha^{\bar{i}\bar{j}}$. 
     This would define a complex family of the BRST transformations $\delta_t$ which are 
      nilpotent $\delta_t^2=0$
      off-shell and the action (\ref{gaugefixedjsklextra}) is invariant under $\delta_t$.
      
      We can repeat the argument from the previous subsection. Using the localization with respect 
       to $\delta_t$ for any $t$  (including zero) and the integration over $\eta$ we arrive
         at the conclusion that the path integral is localized on the constant maps. Thus again 
          the correlator  (\ref{rewiew993030202020})  of local observables is full quantum result. 
       
  The example of holomorphic Poisson structure is provided by the hyperK\"ahler manifold
   which admits a holomorphic symplectic structure with respect to appropriate 
     complex structure.  Therefore the A-model on hyperK\"ahler manifold can be localized 
      to constant maps and the semi-classical result is exact.  However our results are applicable for 
     the   wide class of Poisson holomorphic manifold, e.g. the Del Pezzo surfaces, the Poisson Fano varieties,
      $\mathbb{C}P^2$ etc. 
      These examples have attracted recently a lot attention, especially 
        in the context of generalized complex geometry (see \cite{LaurentGengoux:2007kh, gualtieri2} for the general discussion and the examples \cite{hitchindelP, gualtieri3}). 
        
 One may observe that the PSM for a holomorphic Poisson manifold has a striking similarities 
  with the B-model \cite{Pestun:2006rj} defined for the following generalized complex structure
  \beq
   \left ( \begin{array}{cc}
    J & \alpha \\
    0 & -J^t 
    \end{array} \right )~,
  \eeq{GFDHW999}   
   where $\alpha = \alpha^{(2,0)} + \alpha^{(0,2)}$ is the real part of a holomorphic Poisson structure. 
    However  to define the B-model we need a closed pure spinor 
    $$\rho = e^{\alpha^{(2,0)}} {\bf \Omega}~,$$
     where ${\bf \Omega}$ is a closed holomorphic volume form. Indeed this condition gives
      the holomorphic analog of unimodularity. However for the PSM discussed above we need
       a real version of unimodularity of $\alpha$ which is a weaker condition on a real volume form.
         Thus the unimodular deformations of holomorphic 
        Poisson structure cannot be mapped to the corresponding deformations of generalized Calabi-Yau
         structure corresponding to (\ref{GFDHW999}).   Therefore for a given geometrical setup 
          the B-model and PSM are two different models, with the different moduli dependence.  

\section{Conclusions}
\label{end}

In this work we have attempted to study the Poisson sigma model beyond the perturbative 
 expansion. The main lesson is that the quantum theory requires the corresponding Poisson 
  tensor $\alpha$ to be unimodular. We argued this additional property of $\alpha$ in different ways.
   In the BV framework the unimodularity is related to the quantum master equation, which requires
    an additional care in its definition.  Moreover for the specific
      gauge fixing we obtained the unimodularity 
     as from the requirement of the BRST invariance of the zero mode measure. 
 
 Alternatively one can provide a different heuristic argument\footnote{We thank Alberto Cattaneo for sharing this argument with us. Also see \cite{felder1} for the related discussion and another interesting 
  work \cite{Dolgushev:2006ie} on the relation between the deformation quantization and unimodularity.} 
   for the unimodularity of the Poisson 
  tensor coming from the perturbative analysis as in \cite{Cattaneo:1999fm}. In the perturbative
   expansion all integrals are absolutely convergent except those containing tadpole diagrams. 
   One may try to regularize the tadpoles by the point-splitting using the vector field with no
    zeros on $\Sigma$. However such vector does not exists on $S^2$ and thus the tadpoles should 
     be dealt with differently. Since the tadpoles correspond to the bidifferential operators involving the 
      divergence of Poisson tensor then the unimodularity is the way to eliminate them. 
      
    The unimodulary of Poisson tensor reformulated in terms of pure spinors allows us to 
     treat the PSM exactly in the same fashion as A- and 
      B-models \cite{Hori:2003ic} together with their generalized 
      complex relatives \cite{Kapustin:2003sg, Kapustin:2004gv, Li:2005tz, Pestun:2006rj}. Indeed the Poisson 
       structure defines a real analog of generalized complex structure and the unimodulary of 
       $\alpha$ is a real analog of generalized Calabi-Yau condition. We believe that it is important 
        that all these models can be treated uniformly and there is intricate interrelation between 
         all these models.  
 
 There are several open questions we would like to address in future, in particular
   the generalization  the construction of Frobenius manifolds from \cite{BK} and \cite{manin}
    for the case when the $\d\bar{\d}$-lemma fails, as in a generic Poisson case. Also 
     we plan to use further the localization for PSM along the lines presented 
      in Section \ref{gauge}.  There is an indication that the Gromov-Witten story can be generalized
       for PSM defined over the generalized K\"ahler manifold.  Furthermore it would be interesting 
        to develop the present analysis for PMS for the higher genus surfaces. 

\bigskip\bigskip

\noindent{\bf\Large Acknowledgement}:
\bigskip

\noindent We are grateful to Alberto Cattaneo, Gil Cavalcanti, 
 Andrei Losev, Vasily Pestun, Gabriele Vezzosi and Roberto Zucchini
  for the discussions.
 We thank Alberto Cattaneo, Yvette Kosmann-Schwarzbach and Vasily Pestun for reading and 
  commenting on the manuscript.  
   We thank the referee for the comments and suggestions. 
  We thank  the Erwin Schr\"odinger International Institute for Mathematical Physics for the hospitality. 
 M.Z. thanks INFN Sezione di Firenze  and Universit\`a di Firenze 
  where part of this work was carried out. 
 The research of M.Z. was
supported by VR-grant 621-2004-3177. 

\appendix
\Section{The multivector calculus}
\label{a:PL}

Through out the Appendices A and B we consider mainly the case of compact manifold $M$. 
 This condition can be relaxed if we require the appropriate integrals to be defined and 
  the integration by parts should work without any boundary contributions. 

In this Appendix we review the relevant structures on the multivector fields $\Gamma(\wedge^\bullet TM)$ over a smooth manifold $M$.  For further details the reader may consult the textbook by Vaisman,
  \cite{vaisman}. 

 The Lie bracket on the vector fields can be extended to a bracket on the multivectors. This bracket is 
  called the Schouten bracket.  In local coordinates the multivector fields $P$ and $Q$ are written as
$$ P=P^{{\mu_1}...{\mu_p}}\d_{\mu_1}\wedge...\wedge\d_{\mu_p}$$
$$ Q = Q^{{\mu_1}...{\mu_q}} \d_{\mu_1}\wedge...\wedge\d_{\mu_q}$$
 and their Schouten bracket is defined by the following expression\footnote{Our definition differs by 
  the overall factor $(-1)^{p-1}$ compared to the one in  \cite{vaisman}.} 
  \beq
 [P, Q]_s = \left ( 
  p\, P^{{\mu_1}...{\mu_{p-1}}\rho} \d_\rho Q^{{\mu_p}...{\mu_{q+p-1}}}-q\, \d_\rho P^{{\mu_1}...{\mu_p}} Q^{\rho {\mu_{p+1}}...{\mu_{q+p-1}}} \right ) \d_{\mu_1}\wedge ...
  \wedge \d_{\mu_{q+p-1}}~.
\eeq{defshcoutenbr}
The algebra $(\Gamma (\wedge^\bullet TM), ~\wedge,~ [~,~]_s)$ is a Gerstenhaber algebra (see the definition \ref{Gerstenhaberalg}). 

If further we specify a volume form $\Omega$ on $M$ and a closed one-form $\lambda$ then we can 
 introduce an operator $D_{\Omega,\lambda}$
 $$ D_{\Omega,\lambda} P = \mathbf{div}_\Omega P + i_\lambda P~,$$
  where $\mathbf{div}$ is a divergence operator defined by $\Omega$ and $i_\lambda$ is a contraction 
   with one-form $\lambda$.  In local coordinates with the volume form written as
   $\Omega= \rho~ dx^1\wedge...\wedge dx^d$ the divergence operator is
   $$ (\mathbf{div}_\Omega P)^{\mu_2 ...\mu_p} =  - p \frac{1}{\rho} \d_{\mu_1}\left (\rho ~P^{\mu_1\mu_2...\mu_d} \right )~.$$
   Equivalently, in coordinate free notation,  the divergence can be written as
   $$ \mathbf{div}_\Omega P = -  *^{-1} d *P~,$$
   where $*P = i_P \Omega $ provides a map  from $\Gamma(\wedge^p TM)$ to differential forms and 
   $d$ is de Rham differential. 
   
    Assuming that $d\lambda=0$ we have $(D_{\Omega,\lambda} )^2P =0$
    and moreover
 \beq
 [P, Q]_s =  (-1)^p D_{\Omega,\lambda}(P \wedge Q) + (-1)^{p+1} (D_{\Omega,\lambda}P) \wedge Q
 - P \wedge D_{\Omega,\lambda} Q~.
\eeq{Laplacedefo}
 Indeed $D_{\Omega,\lambda}$ is most general operator which generates
   the Schouten bracket \cite{pingxu}.  Therefore the algebra 
   $(\Gamma (\wedge^\bullet TM), ~\wedge,~ [~,~]_s,~D_{\Omega,\lambda})$ is 
 a BV algebra (see the definition \ref{BValgebra}). 
 
 \begin{definition}
 The bivector $\alpha \in \Gamma(\wedge^2 TM)$ is called a Poisson structure if it satisfies
 $$[\alpha,\alpha]_s=0~.$$ 
 The manifold with such $\alpha$ is called a Poisson manifold. 
 \end{definition}
  The Poisson structure defines a Lichnerowicz-Poisson differential $d_{LP}$ on multivector fields
  $$d_{LP} P \equiv  [\alpha, P]_s~,~~~~P \in \Gamma(\wedge^\bullet TM)~.$$
   The corresponding cohomology $H^\bullet_{LP}(M)$ is
    called the Lichnerowicz-Poisson cohomology group. 
 
  We assume that $M$ is orientable and thus we can choose a volume form $\Omega$. 
    Then we can study how the Hamiltonian vector fields $X_f = \alpha(df)$, $f\in C^\infty(M)$ 
    act on $\Omega$.  In particular there exists a vector field $\phi_\Omega$ such that
    $$ {\cal L}_{X_f} \Omega = \phi_\Omega (f) \Omega~.$$
     $\phi_\Omega$ is named the modular vector field. Indeed the vector field $\phi_\Omega$ 
      defines a class $[\phi_\Omega] \in H^1_{LP}(M)$. This class is independent of $\Omega$, 
      $$ {\cal L}_{X_f} (e^g \Omega) = \left ( \phi_\Omega + \frac{1}{2} d_{LP} g\right )(f) e^g \Omega~$$
       and $[\phi_\Omega]$ is called the Poisson modular class.
   
 \begin{definition}
 A Poisson manifold $(M, \alpha)$ is called unimodular 
  \cite{weinstainmodular} if $[\phi_\Omega]=0$. In other words
  there exists such $\Omega$ that ${\cal L}_{X_f} \Omega= 0$ for any Hamiltonian vector 
   field $X_f$. We refer to such $\Omega$ as an invariant volume form. 
 \end{definition}
  For a Poisson manifold $(M, \alpha)$ we can introduce a (Koszul-)Brylinski differential $\delta_B$ on the differential forms $\Omega^\bullet(M)$
  $$ \delta_B = i_\alpha d - d i_\alpha~,$$
   where $i_\alpha$  is contraction with a Poisson tensor $\alpha$ and $d$ is 
    de Rham differential \cite{koszul}. 
    \begin{theorem}
     A Poisson manifold $(M, \alpha)$ is unimodular if and only if there exists a volume form $\Omega$
      such that $\delta_B \Omega =0$ or alternatively $D_\Omega \alpha=0$. 
    \end{theorem}
    {\bf Proof:}    We use notation $D_\Omega\equiv D_{\Omega, 0}$. 
     The proof of the theorem follows straightforwardly from the relation $\delta_B \Omega =  - i_{\phi_\Omega} \Omega$. This relation arises from the definition of the modular vector field $\phi_\Omega$
      given above and the following identities
     $$d (i_{X_f} \Omega) = - df \wedge \delta_B \Omega~,~~~~~~
     \phi_\Omega(f) \Omega = df \wedge i_{\phi_\Omega} \Omega~.$$
     Moreover using the definition of $D_\Omega$ 
       the modular vector field  can be also defined using the 
        divergence operator with respect to $\Omega$ as $D_\Omega \alpha = - \phi_\Omega$.    
      For more details and the related discussion the reader may 
      consult \cite{weinstainmodular, kosmann2}. $\Box$
      
     Thus we refer to an unimodular Poisson manifold as a triple $(M, \alpha, \Omega)$, where $\Omega$
      is a volume form which is closed under the Brylinski differential. 
 
 \begin{definition}
For a manifold $M$ with a volume form $\Omega$ we define a trace map over the multivector 
 fields 
 $$ tr_\Omega~:~\Gamma (\wedge^{top} TM) \rightarrow {\mathbb R}$$
  as follows 
  $$ tr_\Omega (P) = \int\limits_M \Omega \wedge i_P \Omega~.$$
 \end{definition}
 
 \begin{theorem}\label{theoremPLinterga} 
 For a Poisson manifold $(M, \alpha)$ with a trace map $tr_\Omega$
  the relation
 $$ tr_\Omega (d_{LP} P \wedge Q) = (-1)^{p+1} tr_\Omega (P \wedge d_{LP} Q)$$
  is satisfied if and only if $(M, \alpha)$ is unimodular and $\Omega$ is invariant volume form. 
 \end{theorem}
 {\bf Proof:}  To prove this statement we use the formulas from Vaisman's textbook \cite{vaisman}. 
 The relation in the theorem is equivalent to the following statement
 $$ \int\limits_M \Omega \wedge i_{(d_{LP} W)} \Omega =0~,~~~~~~ W \in \Gamma(\wedge^{d-1}TM).$$
  For this to hold it would be enough to show that $ \Omega \wedge i_{(d_{LP} W)} \Omega$ is an exact $d$-form. Using  the Lichnerowicz definition of the Schouten bracket   (see the formula (1.16) in \cite{vaisman}) we rewrite
   $$ \Omega \wedge i_{(d_{LP} W)} \Omega = - \Omega \wedge i_W \delta_B \Omega +
    (-1)^{d-1} \Omega \wedge \delta_B (i_W \Omega)~.$$
   Assuming that one-form $i_W \Omega = f dg$ and using the properties of the Brylinski differential
    we recast the two terms in the above expression as follows
    $$- \Omega \wedge i_W \delta_B \Omega = (-1)^{d-1} f {\cal L}_{X_g} \Omega~,$$ 
  $$ (-1)^{d-1} \Omega \wedge \delta_B (fdg)= (-1)^d \{ g, f\} \Omega  =   (-1)^d{\cal L}_{X_g} (f\Omega) 
  + (-1)^{d-1} f {\cal L}_{X_g} \Omega ~.    $$
   To derive the first relation we have used $\delta_B \Omega = - i_{\phi_\Omega}\Omega$.
   If we require that the above forms are exact for any
     $g$ and $f$ then the manifold should be unimodular and $\Omega$
       is invariant volume form. Since any one form can be written as sum of the terms like 
    $fdg$ we can extend our proof for a generic situation.  $\Box$

 We can summarize the relevant properties of an unimodular Poisson manifold in 
  the following theorem.
  \begin{theorem}\label{summarize}
   If $(M, \alpha, \Omega)$ is unimodular Poisson manifold then 
   $(\Gamma (\wedge^\bullet TM), ~\wedge,~ [~,~]_s,~D_{\Omega}, d_{LP})$ is a graded 
    differential BV algebra such that 
   $$ D_\Omega d_{LP} + d_{LP} D_\Omega =0~.$$
   Moreover there exists a trace map $tr_\Omega$ such that
      $$ tr_\Omega (d_{LP} P \wedge Q) = (-1)^{p+1} tr_\Omega (P \wedge d_{LP} Q)~,$$
 $$ tr_\Omega ( D_\Omega P \wedge Q) = (-1)^{p} tr_\Omega (P \wedge D_\Omega Q)~.$$
  \end{theorem}
{\bf Proof:}
 The first part of the theorem has been discussed in \cite{pingxu, kosmann2}. 
  We have  explained   most  of the statements  already. The relation between $d_{LP}$
   and $D_\Omega$ is derived as follows
   $$ D_\Omega d_{LP} P = D_\Omega \left ( D_\Omega (\alpha \wedge P) - \alpha \wedge D_\Omega P
    \right ) = -D_\Omega (\alpha \wedge D_\Omega P) = -d_{LP} D_\Omega P~,$$
   where we use the unimodularity, $D_\Omega \alpha=0$. 
    The property of trace with the respect  to the divergence operator $D_\Omega$ is 
   valid for any manifold with a volume form and is just simple consequence of the Stokes theorem  
    for the differential forms.  
$\Box$

\Section{Poisson geometry and pure spinors}
\label{a:spinors}

In this Appendix we reformulate the previous Appendix in a different language. This allows us
 to put the whole formalism into the wider  context which is related to generalized geometry
  on the sum $TM \oplus T^*M\equiv T\oplus T^*$ of the tangent and contangent bundles. 
   Below we review very briefly 
    the notion of generalized complex structure, generalized Calabi-Yau condition
  and their real analogs. For more details we refer the reader to the reviews  
     \cite{gualtieri1, gualtieri2, Zabzine:2006uz}.   

The sum of tangent and cotangent bundles $T\oplus T^*$ has a natural $O(d,d)$ structure given 
 by the natural pairing 
 $$\langle v +\xi, s + \lambda \rangle = \frac{1}{2} (i_v \lambda + i_s \xi)~,$$
  where we adopt the notation $(v+\xi),~(s+\lambda) \in \Gamma(T\oplus T^*)$.  We are interested 
   in a real (complex) Dirac structure which is defined as a maximally isotropic  subbundle 
   of $T\oplus T^*$ (or $(T\oplus T^*) \otimes \mathbb{C}$)  and this subbundle is involutive with
    respect to the Courant bracket.  The Dirac structure is an example of the Lie algebroid with 
     the bracket originated from the restriction of the Courant bracket. 
    In particular we are interested in the case when tangent plus cotangent bundles (or its complexification) can be decomposed as a sum two real (complex) Dirac 
     structures
      $$T\oplus T^* = L \oplus L^*~,~~~~~~~~(T\oplus T^* )\otimes {\mathbb C} = L \oplus L^*~.$$
      This decomposition gives us a real (complex) bialgebroid.  Furthermore there is the structure 
      a differential Gerstenhaber algebra  \cite{kosmann, bialgebroid}
 $$(\Gamma(\wedge^\bullet L^*), \wedge, \{~,~\}, d_{L})~,$$ 
  where $\{~,~\}$  is the extension of the Lie bracket from $L^*$ to $\wedge^\bullet L^*$ and
   $d_L$ is the Lie algebroid differential.  
  In the complex case it is natural to impose an extra condition, namely
     the dual space $L^*$ is complex conjugate of $L$.  Thus the corresponding 
      bialgebroid is
    $$(T\oplus T^* )\otimes {\mathbb C} = L \oplus \bar{L}~.$$
    This special case corresponds to the notion of generalized complex structure 
     \cite{hitchinCY, gualtieri1}. 
    
 Alternatively the Dirac structures can be described by means of the pure spinor lines. 
We define the action of a section $(v+\xi) \in \Gamma (TM \oplus T^*M)$ on a differential form $\rho \in \Gamma(\wedge^\bullet T^*M)$
$$ (v + \xi) \cdot \rho \equiv i_v \rho + \xi \wedge \rho~,$$
   which corresponds to the action of $Cl(T\oplus T^*)$ on $\wedge^\bullet T^*$.
 Thus the differential forms form a natural representation of $Cl(T\oplus T^*)$.
  Consider the Dirac structure $L$ and define a subbundle $U_0$ of $\wedge^\bullet T^*$ as
   follows 
 $$ L = \{  (v + \xi) \in \Gamma(T\oplus T^*)~, (v + \xi) \cdot U_0 =0 \}~.$$
  We refer to $U_0$ as a pure spinor line.  The Dirac structure $L$ induces the alternative 
   grading on the differential forms
  $$  \wedge^\bullet T^* = \bigoplus\limits_{k=0}^{\dim M} U_k~,~~~~~  
   U_k = ( \wedge^k L^*) \cdot U_0 ~, $$
   where $\cdot$ stands for the extension of $Cl(T\oplus T^*)$ action to $\wedge^\bullet T^*$.
 The property that $L$ is involutive under 
  the Courant bracket is equivalent to the following
  $$ d (\Gamma(U_0)) \subset \Gamma(U_1)~,$$
  where $d$ is de Rham differential. 
 Indeed we can define a Dirac structure through the subbundle $U_0$  of $\wedge^\bullet T^*$ 
  with above properties.  With respect to the alternative grading we can decompose the de Rham differential as follows
  $$d= \bar{\d} + \d~,~~~~~~\Gamma(U_{k-1}) \stackrel{\d}{\leftarrow} \Gamma(U_k) \stackrel{\bar{\d}}{\rightarrow} 
  \Gamma(U_{k+1})~,$$
   such that $\d^2=0$ and $\bar{\d}^2=0$.  We borrow the notation from 
    the generalized complex geometry and in present context bar over $\d$ does not mean the complex 
     conjugation. 

 From now on we assume that the bundle $U_0$ is trivial and there exists a global section,  a
  pure spinor form $\rho$ which defines $L$ completely.  The integrability of $L$ is equivalent to 
   the statement 
   $$ d \rho = (v + \xi) \cdot \rho~,$$
   for some section $(v + \xi) \in \Gamma(L^*)$.  Since for given $L$ 
   the pure spinor $\rho$ is defined non uniquely, namely  for any $f \in C^\infty(M)$ the form 
   $e^f\rho$ is also a pure spinor. Thus there is a cohomology class  $[(v+\xi)] \in H^1(d_L)$,
    which is just proportional to the  modular class of the  Lie algebroid \cite{evensluw}. 
     Thus we arrive to the following theorem.
     \begin{theorem}
    The Dirac structure $L$ admits the description in terms of closed pure spinor if and only if
     the corresponding $U_0$ bundle is trivial and  Lie algebroid $L$ is unimodular.  
  \end{theorem}
  Since $U_0$ is a line bundle then its triviality analyzed differently in the real and complex
   cases. For instance,  in the complex case we have to require the
     trivial first Chern class, $c_1(U_0)=0$.  In generalized complex case 
     $(T\oplus T^* )\otimes {\mathbb C} = L \oplus \bar{L}$ the ability to describe $L$ in terms of 
      a closed pure spinor corresponds to the generalized Calabi-Yau condition, the notion introduced 
       by Hitchin \cite{hitchinCY}.  Thus the generalized Calabi-Yau condition is equivalent 
        to two requirements, $c_1(U_0)=0$ and the unimodularity of Lie algebroid $L$.
 
  From now on we assume that $L$ admits the description in terms of closed pure spinor $\rho$. 
  For $A \in \Gamma(\wedge^\bullet L^*)$ and a closed pure spinor $\rho$ there are 
   the following relations 
  $$ (d_L A) \cdot \rho = \bar{\d}( A\cdot \rho)~,~~~~
      (D A) \cdot \rho = \d( A \cdot \rho)~,$$
       where the last relation can be regarded as the definition of the operator $D$ such that $D^2=0$.
         Indeed $D$ generate the bracket $\{~,~\}$ on $\wedge^\bullet L^*$.
   Therefore one can show that  $(\Gamma(\wedge^\bullet L^*), \wedge, \{~,~\}, D,  d_{L})$ is differential BV-algebra \cite{pingxu, Kapustin:2004gv, Li:2005tz}. In addition the closed pure spinor provides
    the isomorphisms of the cohomologies, $H^\bullet (d_L) \approx H^\bullet (\bar{\d})$ and
     $H^\bullet (D) \approx H^\bullet (\d)$.
  
  There exists an invariant form on spinors which, in the present context, corresponds to the Mukai pairing of the differential forms
  $$ (\rho, \phi) = \sum\limits_{j}(-1)^j (\rho_{2j} \wedge \phi_{n-2j} + \rho_{2j+1} \wedge \phi_{n-2j-1})~,$$
   where $n=\dim M$ and the forms decomposed by the standard degree $\rho =\sum \rho_{i}$, $\phi =
    \sum \phi_i$. 
   We can introduce the trace map as
   $$ tr_\rho (A) =  \int\limits_M (\rho, A\cdot \rho)~,~~~~A\in \Gamma(\wedge^n L^*)~. $$
   We can summarize these observation in the following theorem.
  
\begin{theorem} 
  For a Lie bialgebroid $T \oplus T^* = L \oplus L^*$ 
  with $L$ being a Dirac structure described 
   by the a closed pure spinor $\rho$
   $$(\Gamma(\wedge^\bullet L^*), \wedge, \{~,~\}, D,  d_{L})$$
    is differential BV-algebra and there exists trace map with the following properties
    $$ tr_\rho (d_L A \wedge  B) = (-1)^{|A|+1} tr_\rho (A\wedge d_L B)~,$$
    $$ tr_\rho (D A \wedge B) = (-1)^{|A|} tr_\rho (A \wedge DB)~,$$
     where $A, B$ are sections of $\wedge^\bullet L^*$.
\end{theorem}
{\bf Proof:}
 The proof of this theorem is straightforward and the different elements of the proof are scattered 
  in the literature, see \cite{pingxu, Kapustin:2004gv, Li:2005tz}. 
    Let us sketch the main idea behind the proof.
   For any differential form $\rho \in \Gamma(\wedge^\bullet T^*)$ and 
    any sections $A, B \in \Gamma (T\oplus T^*)$ there is the  following identity
    $$ A \cdot B \cdot d\rho = d(A \cdot B \cdot \rho) + B\cdot d(A \cdot \rho ) - A \cdot d(B\cdot \rho)
     + [A, B]_c \cdot \rho - d \langle A , B \rangle \wedge \rho~,$$ 
  where $[~,~]_c$ is the Courant bracket and $\langle~,~\rangle$ is the natural 
  pairing on $T\oplus T^*$. If we have a Lie bialgebroid   $T \oplus T^* = L \oplus L^*$ 
  with $L$ being a Dirac structure described  by the a closed pure spinor $\rho$ then 
   the above formula implies 
    $$  d(A \cdot B \cdot \rho) + B\cdot d(A \cdot \rho ) - A \cdot d(B\cdot \rho)
     + \{A, B\} \cdot \rho =0 ~,$$ 
     where now $A,B \in \Gamma(L^*)$ and $\{~,~\}$ is a Lie bracket on $L^*$, which is a restriction 
      of the Courant bracket to $L^*$. This formula can be extended to the general case 
       when $A, B$ are sections of definite degree in $\Gamma(\wedge^\bullet L^*)$.
        This extension together with the definition 
        $$ (d_L + D)A \cdot \rho = d (A\cdot \rho)~,~~~~~~~ \wedge^k L^* \stackrel{d_L}{\rightarrow}\wedge^{k+1} L^* ~,~~~~~~~\wedge^k L^* \stackrel{D}{\rightarrow}\wedge^{k-1} L^* $$
        we recover that $D$ generates the bracket on $\Gamma(\wedge^\bullet L^*)$ and moreover 
         $\Gamma(\wedge^\bullet L^*)$  is differential BV algebra.  The properties of the trace map 
          can be proven easily using also above properties.
  $\Box$
 
Using this language we now recast the previous definitions in Poisson geometry in a new language. 
 Let us start from the following theorem.

\begin{theorem}
 The manifold $M$ is unimodular Poisson manifold if and only there exists a closed  pure spinor of the  form
 $$ \rho = e^\alpha \Omega = \Omega + i_{\alpha}\Omega + \frac{1}{2} i^2_\alpha \Omega + ...~,$$
  where $\alpha$ is a bivector and $\Omega$ is a volume form.
\end{theorem}
{\bf Proof:} If we have a unimodular Poisson manifold $(M, \alpha, \Omega)$ then we can construct 
 a pure spinor $\rho = e^\alpha \Omega$ which satisfies 
 $$d \rho = \delta_B \Omega + \frac{1}{2} \delta_B (i_\alpha \Omega) + ... = 0~,$$
  since $\delta_B\Omega =0$ and $\delta_B i_\alpha = i_\alpha \delta_B$.  In 
opposite direction we can start from a closed pure spinor $\rho= e^\alpha\Omega$ which 
 defines the following maximally isotropic subbundle of $T\oplus T^*$
 $$L = e^\alpha (T^*) = \{ i_\xi \alpha + \xi~:~ \xi \in \Gamma(T^*)\}~.$$
 Since $\rho$ is closed $L$ is a Dirac structure and  thus $\alpha$ is Poisson structure. Moreover the volume $\Omega$ would be an invariant volume 
   form with respect to the unimodular Poisson structure $\alpha$.
$\Box$

Thus the Poisson structure on $M$ gives the real Lie bialgebroid $T\oplus T^* = e^\alpha(T^*) \oplus T$.
 If the Poisson structure is unimodular then there exists a closed pure spinor $\rho = e^\alpha \Omega$
  and $\Gamma(\wedge^\bullet T)$ is differential BV algebra. 
  Indeed the trace map $tr_\Omega$ defined in the previous appendix coincides with the one defined
   here $tr_\rho$ since the only top form part contributes in $\rho$.

On an unimodular Poisson manifold $(M, \alpha, \Omega)$ with the pure spinor $\rho = e^{\alpha}\Omega$ we can calculate the differentials $\d$ and $\bar{\d}$ associated with the alternative 
 grading on the differential forms
  $$  \wedge^\bullet T^* = \bigoplus\limits_{k=0}^{\dim M} ( \wedge^k T) \cdot e^\alpha \Omega ~. $$
  Indeed in this case we have $\bar{\d} = -\delta_B$ and $\d= d+ \delta_B$, see the following theorem.
\begin{theorem}\label{theorem13}
 For unimodular Poisson manifold $(M,\alpha, \Omega)$ with the closed pure spinor
  $\rho = e^\alpha \Omega$ the following relations hold
$$ (D_\Omega P) \cdot \rho =  - (d + \delta_B) (P\cdot \rho)~,$$
$$ (d_{LP} P) \cdot \rho =  \delta_B ( P \cdot \rho)~.$$
\end{theorem}
{\bf Proof:}
 Let us start from the proof of the first relation. If $\alpha=0$ then this is just a definition of 
  $D_\Omega$ given in the previous appendix. In general case $\alpha \neq 0$ 
 a simple calculation produces the following formula \cite{Evens}
$$  d + \delta_B = e^\alpha d e^{-\alpha}~,$$
 which together with the definition of $D_\Omega$ gives the desired relation.

 Next we prove the second relation in the theorem. 
Using the fact that $D_\Omega$ generates the Schouten bracket and the manifold is unimodular,
 $D_\Omega \alpha =0$ we get
$$ (d_{LP} P) \cdot \rho = \left ( D_\Omega (\alpha \wedge P) - \alpha \wedge D_\Omega P \right )
 \cdot \rho =
 -(d+\delta_B) (i_\alpha i_P \rho) + i_\alpha  (d+\delta_B) (i_P \rho) = \delta_B (i_P \rho)~,$$
 where we used the previously proved relation and the property $i_\alpha \delta_B = \delta_B i_\alpha$.
$\Box$

This theorem implies the isomorphism of certain cohomologies. For any Poisson manifold $(M,\alpha)$
 there are the following isomorphisms
 $$ H^{\bullet}_{dR} (M) \approx H^\bullet (D_\Omega) \approx H^\bullet (d+ \delta_B)~,$$
 while for the unimodular Poisson manifold in addition we have
 $$ H^\bullet_{LP}(M) \approx H^\bullet (\delta_B)~.$$

\eject


\begin{thebibliography}{6666}

\newcommand{\np}{{\em Nucl.\ Phys.\ }}
\newcommand{\pr}{{\em Phys.\ Rev.\ }}
\newcommand{\cmp}{{\em Commun.\ Math.\ Phys.\ }}
\newcommand{\pl}{{\em Phys.\ Lett.\ }}
%
\bibitem{Alexandrov:1995kv}
  M.~Alexandrov, M.~Kontsevich, A.~Schwartz and O.~Zaboronsky,
  ``The Geometry of the master equation and topological quantum field theory,''
  Int.\ J.\ Mod.\ Phys.\ A {\bf 12} (1997) 1405
  [arXiv:hep-th/9502010].
%
\bibitem{BK}
S.~Barannikov and M.~Kontsevich,
``Frobenius Manifolds and Formality of Lie Algebras of
        Polyvector Fields,"
   Internat.\ Math.\ Res.\ Notices {bf 4} (1998) 201
   [arXiv:alg-geom/9710032].     
%
\bibitem{Batalin:1977pb}
  I.~A.~Batalin and G.~A.~Vilkovisky,
  ``Relativistic S Matrix Of Dynamical Systems With Boson And Fermion
  Constraints,''
  Phys.\ Lett.\  B {\bf 69} (1977) 309.
%
\bibitem{Bergamin:2004pn}
  L.~Bergamin, D.~Grumiller, W.~Kummer and D.~V.~Vassilevich,
  ``Classical and quantum integrability of 2D dilaton gravities in  Euclidean
  space,''
  Class.\ Quant.\ Grav.\  {\bf 22} (2005) 1361
  [arXiv:hep-th/0412007].
%
\bibitem{Calvo:2005ww}
  I.~Calvo,
  ``Supersymmetric WZ-Poisson sigma model and twisted generalized complex
  geometry,''
  Lett.\ Math.\ Phys.\  {\bf 77} (2006) 53
  [arXiv:hep-th/0511179].
%
\bibitem{Cattaneo:1999fm}
  A.~S.~Cattaneo and G.~Felder,
  ``A path integral approach to the Kontsevich quantization formula,''
  Commun.\ Math.\ Phys.\  {\bf 212} (2000) 591   
  [arXiv:math.qa/9902090].
%
\bibitem{Cattaneo:2001ys}
  A.~S.~Cattaneo and G.~Felder,
  ``On the AKSZ formulation of the Poisson sigma model,''
  Lett.\ Math.\ Phys.\  {\bf 56} (2001) 163
  [arXiv:math.qa/0102108].
 %
 \bibitem{cattaneobv}
 A.~S.~Cattaneo,
 ``From Topological Field Theory to Deformation Quantization and Reduction,"
   Proceedings of ICM 2006, Vol. III, 339-365 (European Mathematical Society, 2006). 
%
\bibitem{Dolgushev:2006ie}
  V.~Dolgushev,
  ``The Van den Bergh duality and the modular symmetry of a Poisson variety,''
  arXiv:math/0612288.
%
\bibitem{Evens}
S.~Evens and J.-H.~Lu,
``Poisson harmonic forms, Kostant harmonic forms, and the
        $S^1$-equivariant cohomology of $K/T$,"
 Adv.\ Math.\  {\bf 142} (1999) 171
[arXiv:dg-ga/9711019].
%
\bibitem{evensluw}
S.~Evens, J.-H.~Lu and A.~Weinstein,
``Transverse measures, the modular class, and a cohomology pairing for Lie algebroids,"
 Quart. J. Math. Oxford Ser. (2) {\bf 50} (1999), no. 200, 417--436
[arXiv:dg-ga/9610008].
%
\bibitem{felder1}
 G.~Felder and B.~Shoikhet,
 ``Deformation quantization with traces,"
  Lett. Math. Phys. {\bf 53} (2000), no. 1, 75--86
 [arXiv:math.QA/0002057].
%
\bibitem{fiorenza}
D.~Fiorenza,
``An introduction to the Batalin-Vilkovisky formalism,"
Comptes Rendus des Rencontres Mathematiques de Glanon, Edition 2003
[arXiv:math.QA/0402057].
%
\bibitem{Frenkel:2005ku}
  E.~Frenkel and A.~Losev,
  ``Mirror symmetry in two steps: A-I-B,''
  Commun.\ Math.\ Phys.\  {\bf 269} (2007) 39
  [arXiv:hep-th/0505131].
%
\bibitem{gualtieri1}
M.~Gualtieri, ``Generalized complex geometry,'' Oxford University  
DPhil thesis,
arXiv:math.DG/0401221 .
%
\bibitem{gualtieri2}
M.~Gualtieri, ``Generalized complex geometry,'' 
arXiv:math.DG/0703298 .
%
\bibitem{gualtieri3}
M.~Gualtieri,
``Branes on Poisson varieties,"
arXiv:0710.2719 .
%
\bibitem{henneaux}
M.~Henneaux and C.~Teitelboim,
``Quantization of gauge systems,"
Series:Princeton Series in Physics 1992.
%
%
\bibitem{Hirshfeld:2001cm}
  A.~C.~Hirshfeld and T.~Schwarzweller,
  ``The partition function of the linear Poisson-sigma model on arbitrary
  surfaces,''
  arXiv:hep-th/0112086.
%
\bibitem{hitchinCY}
N.~Hitchin,
``Generalized Calabi-Yau manifolds,"
Quart.\ J.\ Math.\ Oxford Ser. {\bf 54} (2003) 281
[arXiv:math.DG/0209099].
%
\bibitem{Hitchin:2005cv}
  N.~Hitchin,
  ``Instantons, Poisson structures and generalized Kaehler geometry,''
  Commun.\ Math.\ Phys.\  {\bf 265} (2006) 131
  [arXiv:math.dg/0503432].
%
\bibitem{hitchindelP}
N.~Hitchin,
``Bihermitian metrics on Del Pezzo surfaces,"
 arXiv:math.DG/0608213.
%
\bibitem{Hori:2003ic}
  K.~Hori, S.~Katz, A.~Klemm, R.~Pandharipande, R.~Thomas, C.~Vafa, R.~Vakil, E.~Zaslow,
  ``Mirror symmetry,''
{\it  Providence, USA: AMS (2003) 929 p}.
%
\bibitem{Ikeda:1993fh}
N.~Ikeda,
``Two-dimensional gravity and nonlinear gauge theory,''
Annals Phys.\  {\bf 235} (1994) 435
[arXiv:hep-th/9312059].
%
\bibitem{Kapustin:2003sg}
  A.~Kapustin,
  ``Topological strings on noncommutative manifolds,''
  Int.\ J.\ Geom.\ Meth.\ Mod.\ Phys.\  {\bf 1} (2004) 49
  [arXiv:hep-th/0310057].
%
\bibitem{Kapustin:2004gv}
  A.~Kapustin and Y.~Li,
  ``Topological sigma-models with H-flux and twisted generalized complex
 manifolds,''
  arXiv:hep-th/0407249.
%
\bibitem{kosmann}
Y.~Kosmann-Schwarzbach,
``Exact Gerstenhaber algebras and Lie bialgebroids,"
Acta \ Appl.\ Math. {\bf 41} (1995) 153-165.
%
\bibitem{kosmann2}
Y.~Kosmann-Schwarzbach,
``Modular vector fields and Batalin-Vilkovisky algebras,"
 Poisson geometry (Warsaw, 1998), 109--129, 
Banach Center Publ., {\bf 51}, Polish Acad. Sci., Warsaw, 2000.
%
\bibitem{kosmann3}
Y.~Kosmann-Schwarzbach and J.~Monterde,
``Divergence operators and odd Poisson brackets,"
Ann. Inst. Fourier (Grenoble) {\bf 52} (2002), no. 2, 419--456
[arXiv:math.QA/0002209].
%
\bibitem{koszul}
J.-L.~Koszul,
``Crochet de Schouten-Nijenhuis et cohomologie,"
The mathematical heritage of \'{E}lie Cartan (Lyon, 1984), 
Ast\'{e}risque 1985, Numero Hors Serie, 257--271. 
%
\bibitem{Krotov:2006th}
  D.~Krotov and A.~Losev,
  ``Quantum field theory as effective BV theory from Chern-Simons,''
  arXiv:hep-th/0603201.
%
\bibitem{Kummer:1996hy}
  W.~Kummer, H.~Liebl and D.~V.~Vassilevich,
  ``Exact path integral quantization of generic 2-D dilaton gravity,''
  Nucl.\ Phys.\  B {\bf 493} (1997) 491
  [arXiv:gr-qc/9612012].
%
\bibitem{LaurentGengoux:2007kh}
  C.~Laurent-Gengoux, M.~Stienon and P.~Xu,
  ``Holomorphic Poisson Structures and Groupoids,''
  arXiv:0707.4253 [math.DG].
%
\bibitem{Li:2005tz}
  Y.~Li,
  ``On deformations of generalized complex structures: The generalized
  Calabi-Yau case,''
  arXiv:hep-th/0508030.
%
\bibitem{bialgebroid}
Z.~J.~Lu, A.~Weinstein and and P.~Xu,
``Manin Triples for Lie Bialgebroids,"
J.\ Diff.\ Geom.\ {\bf 45} (1997) 547
[arXiv:dg-ga/9508013].
%
\bibitem{Lyakhovich:2004kr}
  S.~L.~Lyakhovich and A.~A.~Sharapov,
  ``Characteristic classes of gauge systems,''
  Nucl.\ Phys.\  B {\bf 703} (2004) 419
  [arXiv:hep-th/0407113].
%
\bibitem{Lyakhovich:2002kc}
  S.~Lyakhovich and M.~Zabzine,
  ``Poisson geometry of sigma models with extended supersymmetry,''
  Phys.\ Lett.\ B {\bf 548} (2002) 243
  [arXiv:hep-th/0210043].
%
\bibitem{manin}
Yu.~I.~Manin,
``Three constructions of Frobenius manifolds: a comparative study,"
In {\it Surveys in differential geometry}, 497-554,
 Surv. Differ. Geom., VII, Int. Press, Somerville, MA, 2000
[arXiv:math.QA/9801006].
%
\bibitem{maninbook}
Yu.~I.~Manin,
 ``Frobenius manifolds, quantum cohomology, and moduli spaces,"
  {\it American Mathematical Society Colloquium Publications}, {\bf 47},
   American Mathematical Society, Providence, RI, 1999. xiv+303 pp.
%
\bibitem{Mnev:2006ch}
  P.~Mnev,
  ``Notes on simplicial BF theory,''
  arXiv:hep-th/0610326.
%
\bibitem{Pestun:2006rj}
  V.~Pestun,
  ``Topological strings in generalized complex space,''
  Adv.\ Theor.\ Math.\ Phys.\  {\bf 11} (2007) 399
  [arXiv:hep-th/0603145].
%
\bibitem{roytenberg1}
D.~Roytenberg,
``On the structure of graded symplectic supermanifolds and
        Courant algebroids,"
        Quantization, Poisson Brackets and Beyond, Theodore Voronov (ed.), Contemp. Math., Vol. {\bf 315}, Amer. Math. Soc., Providence, RI, 2002
        [arXiv:math.SG/0203110].
%
\bibitem{Schaller:1994es}
P.~Schaller and T.~Strobl,
``Poisson structure induced (topological) field theories,''
Mod.\ Phys.\ Lett.\ A {\bf 9} (1994) 3129
[arXiv:hep-th/9405110].
%
\bibitem{Schwarz:1992nx}
  A.~S.~Schwarz,
  ``Geometry of Batalin-Vilkovisky quantization,''
  Commun.\ Math.\ Phys.\  {\bf 155} (1993) 249
  [arXiv:hep-th/9205088].
%
\bibitem{vaisman}
 I.~Vaisman, 
 âLectures on the geometry of Poisson manifolds,â Progress in Mathemat- 
ics, {\bf 118}. Birkh Ìauser Verlag, Basel, 1994. viii+205 pp.
%
\bibitem{weinstainmodular}
A.~Weinstein, ``The modular automorphism group of Poisson manifolds,"
J.\ Geom.\ Phys.\ {\bf 23} (1997) 379.
%
\bibitem{Witten:1988xj}
  E.~Witten,
  ``Topological Sigma Models,''
  Commun.\ Math.\ Phys.\  {\bf 118} (1988) 411.
%
\bibitem{Witten:1991zz}
  E.~Witten,
  ``Mirror manifolds and topological field theory,''
  arXiv:hep-th/9112056.
  %
  \bibitem{pingxu}
  P.~Xu,
  ``Gerstenhaber algebras and BV-algebras in Poisson geometry,"
  Comm.\ Math.\ Phys.\  {\bf 200} (1999) 545
  [arXiv:dg-ga/9703001].
  %
\bibitem{Zabzine:2006uz}
  M.~Zabzine,
  ``Lectures on generalized complex geometry and supersymmetry,''
  Archivum mathematicum (supplement) {\bf 42} (2006) 119-146
  [arXiv:hep-th/0605148].
  %
\bibitem{Zucchini:2004ta}
  R.~Zucchini,
  ``A sigma model field theoretic realization of Hitchin's generalized  complex
  geometry,''
  JHEP {\bf 0411} (2004) 045
  [arXiv:hep-th/0409181].
 %
\bibitem{Zucchini:2005cq}
  R.~Zucchini,
  ``A topological sigma model of biKaehler geometry,''
  JHEP {\bf 0601} (2006) 041
  [arXiv:hep-th/0511144].



\end{thebibliography}
\end{document}